\documentclass[aps,preprint,showpacs,superscriptaddress,groupedaddress,nofootinbib]{revtex4}  
\usepackage{amssymb,amsmath,bm,graphicx}

\newcommand{\gev}{{\rm \, GeV}}

\begin{document}

\title{Effects of local event-by-event conservation laws in ultra-relativistic heavy ion collisions at the particlization}

\author{Dmytro Oliinychenko}
\affiliation{Lawrence Berkeley National Laboratory, 1 Cyclotron Rd, Berkeley, CA 94720, US}
\author{Shuzhe Shi}
\affiliation{Department of Physics, McGill University, Montreal, QC H3A 2T8, Canada}
\author{Volker Koch}{1}
\affiliation{Lawrence Berkeley National Laboratory, 1 Cyclotron Rd, Berkeley, CA 94720, US}

\begin{abstract}
  Many simulations of relativistic heavy ion collisions involve the switching from relativistic hydrodynamics to kinetic particle transport. This switching entails the sampling of particles from the distribution of energy, momentum and conserved currents provided by hydrodynamics. Usually this sampling ensures the conservation of these quantities only on the average, i.e. the conserved quantities may actually fluctuate among the sampled particle configurations and only their averages over many such configurations agree with their values from hydrodynamics. Here we  apply a recently invented method \cite{Oliinychenko:2019zfk} to ensure  conservation laws for each sampled configuration in spatially compact regions (patches) and study their effects: from the well-known (micro-)canonical suppression of means and variances to little studied (micro-)canonical correlations and higher order fluctuations. Most of these effects are sensitive to the patch size. Many of them do not disappear even in the thermodynamic limit, when the patch size goes to infinity. The developed method is essential for particlization of stochastic hydrodynamics. It is useful for studying the chiral magnetic effect, small systems, and in general for  fluctuation and correlation observables.
\end{abstract}

\pacs{Relativistic heavy-ion collisions, Particle correlations and fluctuations}
\maketitle

\section{Introduction}

One of the major goals of relativistic heavy ion collision experiments is to study a transition from a hadron gas to the quark-gluon plasma. The RHIC Beam Energy Scan experimental program is devoted to searching for the critical point of this transition by lowering collision energy from 200 GeV per nucleon pair down to 7 GeV. Future experiments at NICA and FAIR, which also have this search as one of their goals, are currently under construction. Multiple phenomenological models predict the critical point, but its location on the phase diagram varies considerably from model to model, and scenarios, where the critical point exists but is not experimentally accessible, are not excluded. The most promising experimental signatures of the critical point seem to be enhanced fluctuations. Therefore, considerable attention is devoted to correlation and fluctuation observables, such as proton, net-proton, net-charge, and kaon cumulants~\cite{Stephanov:2008qz,Luo:2015ewa, Adamczyk:2017wsl,Adam:2020unf} and correlations \cite{Adam:2019xmk}, fluctuations of various particle ratios~\cite{Abdelwahab:2014yha}, transverse momentum correlations~\cite{Adam:2019rsf}, and charge balance functions~\cite{Adamczyk:2015yga} (for a recent review see \cite{Bzdak:2019pkr}). Other promising observables include light nuclei production, which may be related to fluctuations through coalescence \cite{Sun:2018jhg}.

Understanding these observables and linking them to a possible critical point requires dynamical modelling of heavy ion collisions, which includes treatment of fluctuations and correlations. Using a transport approach is a possible way (for a transport code including both partons and hadrons see e.g. PHSD \cite{Cassing:2008sv}), but it inevitably results in large theoretical uncertainties related to hadronisation, because the exact mechanism of hadronization is not known. Moreover, PHSD uses a test-particle method, which artificially reduces correlations, hence the recent effort to develop a PHQMD approach \cite{Aichelin:2019tnk} free of this limitation. An alternative way are the hydrodynamic and/or hybrid (hydrodynamic + transport) simulations, where hadronization is encoded in the equation of state (EoS). The latter is parameterized -- it is not know from first principles at large baryon density -- and the parameters can be adjusted to fit measured particle yields, spectra, flow, correlations, and other observables. The effects of the critical point enter the EoS, but this is insufficient to model the vicinity of the critical point. In addition, the slow critical modes have to be explicitly taken into account in the hydrodynamic equations. This is done in the fluctuating hydrodynamics extended by stochastic terms directly~\cite{Kapusta:2011gt,Murase:2013tma,Kumar:2013twa,Nahrgang:2018afz} or coupled to a non-equilibrium field with a stochastic noise~\cite{Nahrgang:2011mg,Herold:2014zoa}. A deterministic approach to treat second order correlations and fluctuations (``hydro+'') is also available \cite{Stephanov:2017ghc}.

Hybrid approaches involving hydrodynamics (fluctuating or not) need a particle sampler to convert hydrodynamic fields to particles that subsequently evolve according to kinetic equations including collision and possibly mean field dynamics. To study correlations or fluctuations, a key requirement for such a sampler is that it  conserves energy-momentum and charges in every event.  Otherwise correlations originating from conservation laws are lost and fluctuations are uncontrollably enhanced \cite{Steinheimer:2017dpb}. In other words, the sampler should be a local microcanonical sampler. Let us explain this term in detail.

Consider an ensemble of particlization hypersurfaces $H_i$, $i=1,N$ obtained from hydrodynamic simulations. For example, the hypersurfaces $H_i$ can be from simulations obtained with different initial conditions. Or $H_i$ may represent ensembles of fluctuating hydrodynamics, where critical fluctuations are explicitly embedded. Usually these are hypersurfaces of constant time, constant energy density, constant temperature, or constant Knudsen number. Suppose that we construct sets of particles from each $H_i$ (we refer to  this as ``performing particlization'') multiple times. Thus we obtain sets of particles $P_{ij}$, where $j \in \{1,\ldots N_{samples}\}$ for each $H_i$.  If the transformation $H_i \to P_{ij}$ is performed in such a way that conservation of energy, momenta, and discrete charges is only fulfilled on average by $j$ (meaning for example, that $\frac{1}{N_{samples}}\sum_{j} Q_j \to Q_{hydro}$ as $N_{samples} \to \infty$, but $Q_j \ne Q_{hydro}$), we call this ``grand-canonical sampling''. If conservation laws are fulfilled for every $j$, meaning $Q_j = Q_{hydro}$, then we call it ``microcanonical sampling'', or event-by-event conservation laws. If this is the case not only for the entire simulation region, but also for smaller space-time regions, we call it ``local microcanonical sampling''. In case of one sample per hypersurface, by construction, local microcanonical sampling preserves fluctuations of conserved quantities over $H_i$ and transfers them to $P_{ij}$ without changes. In contrast, the grand-canonical sampling generates additional fluctuations by allowing conserved quantities of generated particles to differ from those of hydrodynamic events \cite{Steinheimer:2017dpb}. In this paper we are concerned with preserving correlations and fluctuations and, therefore, wish to adopt local microcanonical sampling. In our recent work \cite{Oliinychenko:2019zfk} we invented, described, and tested a method to do it. Several previous attempts fulfilled only some conservation laws (but never all of them), and most of them were generating ad hoc distributions different from the actual microcanonical one, see \cite{Schwarz:2017bdg} for an overview. In this work we apply the method of \cite{Oliinychenko:2019zfk} accounting for energy-momentum, baryon number, strangeness, and charge conservation microcanonically. We do not account for angular momentum or parity conservation, although in principle they can be included too.

We have previously pointed that the sampling should be ``local''. However, the degree of localness is not immediately obvious. At first glance it may seem, that the more local, the better. Because the numerical solution of hydrodynamic equations is often obtained on a discrete space-time grid, it seems easy and practical to enforce conservation laws in every cell of this computational grid. However, in typical simulations of heavy ion collisions these cells are so small that the average number of particles per cell is below one. This problem is not typical for other fields, but one can always obtain a similar situation by choosing a sufficiently  fine computational grid. Another case, where a similar problem can emerge, is a simulation of a very dilute solution, where a number of fluid molecules per cell is large, but average number of dissolved molecules per cell is below 1. Introducing fractional particles is a viable option to approach this problem \cite{Steinheimer:2017dpb}, but the subsequent treatment of these fractional particles in the transport is challenging. Here we explore an alternative way: we define regions, where conservation laws will be fulfilled.

It follows from the considerations above that the scale $b$, on which local conservation laws have to be enforced, cannot be arbitrarily small. It has to be large enough to include the hydrodynamic scale, in other words it should be larger than the mean free path. Also, a patch of size $b$ should contain much more than one particle on average, so $\rho b^3 \gg 1$, where $\rho$ is particle density. On the other hand, $b$ should be smaller than the typical length of correlation one wishes to study. We explore it further by dividing a hypersurface into ``patches'', where conservation laws are enforced, and varying the size of the patch.

Our goal in this paper is to test a local microcanonical sampler, systematically trying different patch sizes and different ways to partition a hypersurface. The methodology is discussed in Sec.~\ref{sec:Methodology}, which comprises the splitting hypersurface into patches where conservation laws are enforced (Sec.~\ref{sec:splitting_patches}), sampling algorithm in a patch (Sec.~\ref{sec:sampling}), discussion of its convergence and runtime (Sec.~\ref{sec:runtime}), and the issue of negative contributions which is known to plague grand-canonical samplers (Sec.~\ref{sec:negative_contributions}). After extensive testing described in Appendices~\ref{appendix:massless_microcanonical} and \ref{appendix:Begun_microcanonical}, we proceed to apply the sampler to a realistic hypersurface in Sec.~\ref{sec:results}, where we compute means, correlations, and higher order fluctuations of particle multiplicities and conserved quantities within a rapidity cut; and also study spectra and flow. All these are done as a function of the patch size. Summary, discussion, and outlook follow in Sec.~\ref{sec:summary}. We made the code for partitioning the hypersurface and microcanonical sampling publicly available at \cite{Oliinychenko:code}.

\section{Methodology} \label{sec:Methodology}

In practice the particlization hypersurface is given as a list of cells with space-time coordinates $x^{\mu}_j$, velocities $u^{\mu}_j$, temperatures $T_j$, chemical potentials $\mu_{B_j},\mu_{S_j},\mu_{Q_j}$, and normal 4-vectors $d\sigma^{\mu}_j$. Our task here is twofold: (i) partition hypersurface into patches, where local conservation laws are enforced, and (ii) sample particles within every patch, while accounting for variations of above quantities within a patch. The latter is crucial, if one wants to ensure that observables sensitive to these variations, such as higher order azimuthal asymmetries, are not smeared out. Tasks (i) and (ii) are independent, therefore we describe them separately.

\subsection{Splitting the  hypersurface into patches} \label{sec:splitting_patches}

Above we have already started introducing a spatial scale $b$, over which  conservation laws should be enforced. There is a number of questions to be addressed regarding this scale.
What is its physical meaning? How big or small should it be? How to partition a hypersurface into patches, given that such partitioning is not unique even if the patch size is fixed? Which observables depend on the choice of patch size and the way of partitioning, and how significant are these dependencies? These are the questions we will discuss in this section.

As already pointed out, the spatial extent of the patch should be neither too small nor too large. By Lorentz-boosting the hypersurface one can see that the same is true for the time extent of the patch, therefore we further call $b$ a \textit{space-time} size. Already from a condition $\rho b^3 \gg 1$ it is clear that the space-time size of all patches cannot be the same, because the local particle density varies. If one chooses to have patches of the same space-time size, then for certain particlization hypersurfaces (to larger extent for isochronous, to lesser extent for iso-energy-density) some patches will contain many particles on average~\footnote{\label{note:average}Here ``average number of particles'' is specifically the grand-canonical mean computed from hydrodynamic variables, as given by Eq. (\ref{eq:cf_integrals}).}, and some will contain less than one particle. Furthermore, particles within patches of the same spatial size but different density would have different mean free paths. In addition to breaking the $\rho b^3 \gg 1$ condition, such a situation is undesirable for our study of microcanonical effects, because we prefer to have the patch size as an interpretable and uniform control parameter of microcanonical effects. Indeed, given the same $b$ for all patches, patches with larger density are less sensitive to microcanonical effects, and patches with smaller density are more sensitive. Therefore we suggest to control the patch size by its rest frame energy. This ensures that our requirements for patches are always fulfilled by construction. An alternative possibility is to use average number of particles per patch for this purpose. We tried this and obtained similar results to those presented here. For the rest of the paper, our patches are formed by combining the nearest hydrodynamic computational cells in space-time until the required rest frame energy $E_{patch}$ is reached. The energy $E_{patch}$ is a parameter that uniformly controls the size of microcanonical effects.

The parameter $E_{patch}$ is not just a technical parameter, it has a clear physical meaning. Suppose that the quark-gluon fluid turns from a continuous stream into separate droplets. Then $E_{patch}$ is the rest frame energy of one droplet, assuming of course that the droplets have the same size. Here we purposefully adopt this assumption to obtain systematic results as a function of $E_{patch}$. However, physics-wise droplets can be of different sizes.

We would like to notice, that scenarios with droplet formation are usually overlooked by the hydrodynamic simulations, where density at the boundaries drops to zero continuously, because the sharp surface and surface tension are not included. If they are accounted for, it may lead to an onset of a well-known Plateau-Rayleigh instability, where a continuous flow of fluid turns into droplets if the ratio of the kinetic energy to the surface energy (the Weber number) is large enough. This phenomenon is ubiquitous and can be observed, for example, in the usual flow of water from a faucet. It is conceivable, that a similar separation into droplets occurs in heavy ion collisions. Moreover, the Plateau-Raleigh instability is not the only possibility to create droplets. They could also be formed as a consequence of cavitation or due to spinodal instabilities \cite{Randrup:2003mu}. Regardless of the mechanism, if the droplets are formed, we show that it has observable consequences, such as suppressed number of particles at high $p_T$, and enhanced $v_2$ at high $p_T$, and these observables depend on $E_{patch}$.

While $E_{patch}$ has a physical meaning and will be further studied as a physical parameter, it is not sufficient to uniquely define the partitioning into patches. The  partitioning also depends on the algorithm, which we will next  describe: We start by choosing an unclustered (not belonging to any patch) cell and add the closest unclustered cells until the total rest frame energy reaches $E_{patch}$. Then the selected cells form a patch and the procedure is repeated until no unclustered cells remain. In this algorithm there are two choices to be made: (i) how to select the initial cell, (ii) how to define the distance to look for closest cells. An additionally uncertainty arises from the fact that in a given patch the conserved charges, such as baryon number,strangeness, and electric charge are most likely non-integer. For microcanonical sampling, however, they have to be integer numbers, as it generates particles with integer charge. Therefore there is an additional  algorithmic choice (iii) of how to assign integer conserved charges to the patches. Next we discuss the choices (i-iii) and explore, how much they influence results. For the choices (i-ii) the following combinations have been tested:
\begin{enumerate}
    \item starting from a cell with minimal time, clustering by distance $\Delta t^2 + \Delta r^2$
    \item starting from a cell with maximal spatial rapidity $\eta = \sqrt{t^2 - z^2}$, clustering by distance $\Delta t^2 + \Delta r^2$
    \item starting from a cell with maximal spatial rapidity $\eta$, clustering by distance in spatial rapidity $\Delta \eta$; this choice has the  advantage of Lorentz-invariance so that the partitioning remains the same for a boosted hypersurface
    \item starting from a cell with maximal energy, clustering by distance $\Delta r^2/d_0^2 + (\Delta T / \sigma_T)^2 + (\Delta \mu_B / \sigma_{\mu_B})^2$, where $\sigma_T$ and $\sigma_{\mu_B}$ are the scaled variances of temperature and baryochemical potential $\mu_B$ over the hypersurface, and  $d_0 = 2$ fm; the idea of this choice is to form patches around hot spots and reduce variations in temperature and baryochemical potential within a patch.
\end{enumerate}

As already mentioned, after the patches are formed in this way, their net charges $B_k$, $S_k$, $Q_k$, $k=1,N_{patch}$ are not necessarily integer numbers. While this is not wrong by itself, the microcanonical sampling requires that the net charges of the patch are integers, because sampled particles always have integer quantum numbers. A simple approach of rounding net charges to a nearest integer may violate global conservation laws. To illustrate this imagine a hypersurface with a net baryon number 50, split into 200 patches each having baryon number 0.4; after rounding procedure every patch will have baryon number 0, therefore the whole hypersurface will have baryon number 0. Certainly, such scenario is undesirable. We would like to preserve the correct conserved quantities (energy-momentum, net baryon number, net strangeness, net charge) of the entire hypersurface: energy-momentum $P_{tot}$, baryon number $B_{tot}$, strangeness $S_{tot}$, and electric charge $Q_{tot}$.
The total conserved charges of the entire system, $B_{tot}$, $S_{tot}$, and $Q_{tot}$, are integers and are conserved during the hydrodynamic evolution~\footnote{It may be that due to numerics or due to the construction of the initial state that the total charges of the whole hypersurface are not integers. In this case we round them to the nearest integer.}. Given a hypersurface where the particlization, i.e. the transition from hydrodynamic fields to particles takes place, the total charges are related to the phase-space density obtained from the hydrodynamic fields by   
\begin{eqnarray} \label{Eq:cons_laws}
\begin{pmatrix}
P_{tot}^{\mu} \\
B_{tot} \\
S_{tot} \\
Q_{tot}
\end{pmatrix} = \sum_{\substack{ \mathrm{cells,}\\i}} \int
\begin{pmatrix}
p^{\mu}_i\\
B_i \\
S_i \\
Q_i
\end{pmatrix}
\frac{p^{\nu} d\sigma_{\nu}}{p^0}
\mathit{f}_i(p^{\alpha}u_{\alpha}, T, \mu_i) \frac{ g_i d^3p}{(2 \pi \hbar)^3}.
\end{eqnarray}
Here the index $i$ runs over all hadronic species with degeneracy $g_i$, $\mu_i = \mu_B B_i + \mu_S S_i + \mu_{Q} Q_i$ is the chemical potential, $\mathit{f}_i$ is the distribution function (in our case it is always J\"uttner distribution), $d \sigma_{\nu}$ denotes the normal 4-vector to the hypersurface cell which is a relativistic analog of volume (see \cite{Huovinen:2012is} for a detailed definition), $T$ is the temperature of this element. In the above formula we sum over all cells of the hypersurface to obtain the total charges. The conserved charges in a given patch $k$ are then given by the same expression where we only sum over the cells in this patch. Consequently, the total charges are the given by the sum of the charges in all patches, $\sum_k B_k = B_{tot}$, $\sum_k S_k = S_{tot}$, and $\sum_k Q_k = Q_{tot}$. However, as discussed above, there is not reason that  the charges in a given patch are integer. To achieve this we need to make additional algorithmic assumptions/choices. As we do it for every charge independently, let us discuss only the baryon number. The non-integer remainders in every patch are $w_k = B_k - \lfloor B_k \rfloor$ and they satisfy $\sum_k w_k = B_{tot} - \sum_k \lfloor B_k \rfloor \equiv B$ and $0 < w_k < 1$. Here $\lfloor x \rfloor$ denotes the floor of $x$. In every patch we need to turn the non-integer remainder $w_k$ either into $\sigma_k = 0$ or $\sigma_k = 1$, while preserving their sum. We propose to do this stochastically with the probability of having  $\sigma_k = 1$ being  proportional to $w_k$. A formal expression for such combined probability is
\begin{eqnarray}
  w(\{\sigma_1, \sigma_2, \dots, \sigma_{N_{patch}}\}) \sim \prod_{k=1}^{N_{patch}} w_k^{\sigma_k} \times \delta\left(\sum_k \sigma_k - B\right)
\end{eqnarray}

\begin{figure*}
    \centering
    \includegraphics[width=\textwidth]{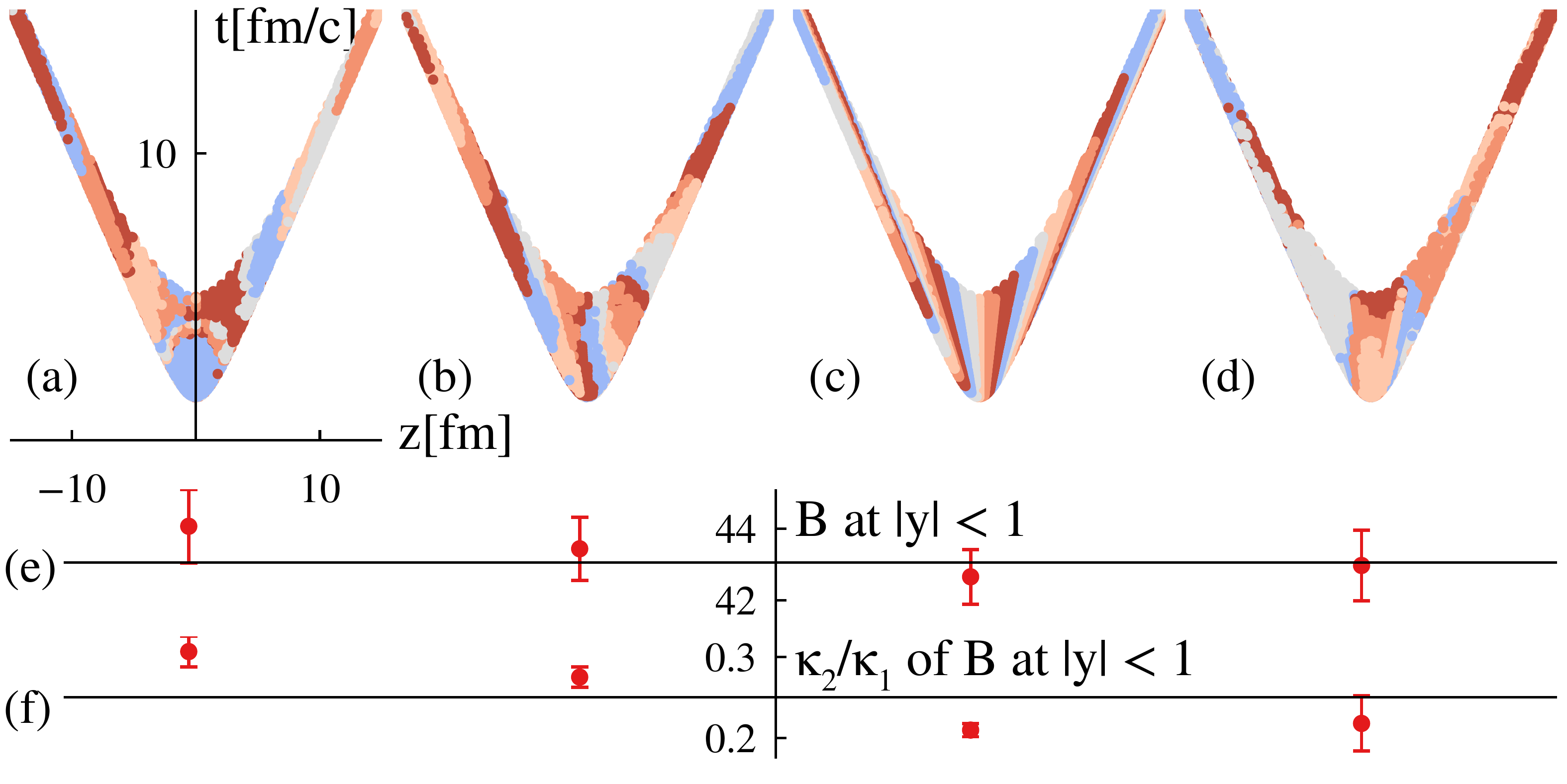}
    \caption{Illustration of the partitioning into patches, and how it influences observables. Panels (a-d) show different ways of partitioning (see text): (a) starting with $t_{min}$, distance $\Delta t^2 + \Delta r^2$; (b) starting with $\eta_{max}$, distance $\Delta t^2 + \Delta r^2$; (c) starting with $\eta_{max}$, distance $\Delta \eta$; (d) starting with $E_{max}$, distance $\Delta r^2/d_0^2 + (\Delta T / \sigma_T)^2 + (\Delta \mu_B / \sigma_{\mu_B})^2$. Panels (e) and (f) show how total baryon number at midrapidity and its variance change depend on the algorithm. The error bars are systematic errors due to requiring integer charges within patches (see text). The hypersurface is the same realistic hypersurface from Au+Au collisions at 19.6 GeV that is used for physics results.}
    \label{fig:different_splitting}
\end{figure*}

In other words, this is a weighted permutation of $B$ ones and $N_{patch} - B$ zeros, with weights proportional to $w_k$. This distribution is generated using a Metropolis walk (the general description of Metropolis algorithm is given further). One step of such walk proposes to exchange a zero at random position $k_1$ with a one at random position $k_2$. This is accepted with probability $min(1, w_{k_2}/w_{k_1})$. After sufficiently many steps we arrive at a sample from the required distribution. This last step completes the separation of the hypersurface into patches: Each patch has a set of cells, that belong to it, it has integer total charges, and its total rest frame energy is close to $E_{patch}$. The remaining question is, how much our algorithmic choices influence physical observables.

This question is addressed throughout the paper by showing all results for two ways of splitting:
maximal $\eta$ first cell and distance in $\eta$ (panel (c) of Fig. \ref{fig:different_splitting}) and the largest energy cell and distance $\Delta r^2/d_0^2 + (\Delta T / \sigma_T)^2 + (\Delta \mu_B / \sigma_{\mu_B})^2$ (panel (d) of Fig. \ref{fig:different_splitting}). Here we additionally explore all the ways of splitting described above for the variables that turned out to be one of the most sensitive to splitting algorithm --- the baryon number at midrapidity and its fluctuations. We use the same hypersurface that is also used for the results subsequently discussed. For every way of splitting described above we produce $10^3$ samples 20 times. For each of these 20 times there is a new assignment of integer quantum numbers to the patches. For each time we compute mean and scaled variance of 
the baryon number within midrapidity, $|y| < 1$. Then we show the means over these 20 times for the baryon number in panel (e) and for its scaled variance in panel (f) of Fig.~\ref{fig:different_splitting}. The variances of these quantities over the 20 times are shown as error bars. Therefore, the error bars represent a systematic uncertainty due to the assignment of integer quantum numbers to the patches. The difference between points in panels (e) and (f) from one splitting method to another is the systematic uncertainty due to the method of splitting hypersurface into patches.

 As seen in panel (f), the assignment of integer baryon numbers within patch matters less for the scaled variance of the baryon number at midrapidity, likely because it is an intensive quantity. However, the scaled variance exhibits a clear sensitivity to the method of splitting. This is understandable, because on our hypersurface the mean baryon number is mainly a function of rapidity $\eta$. Therefore, if one splits the hypersurface by $\eta$ as shown in panel (c), the scaled variance of the baryon number at midrapidity is smaller. If one splits the hypersurface as shown in panel (a) of Fig. \ref{fig:different_splitting}), one patch typically comprises a larger rapidity window and the scaled variance of the baryon number at midrapidity is larger. As a summary, the influence of the patch splitting algorithm on the physical observables is not negligible and should be controlled carefully. We subsequently do it by repeating all our findings for two different splitting methods. The difference between the two should be understood as a systematic uncertainty of our method.

\subsection{Sampling particles in a patch with event-by-event conservation laws} \label{sec:sampling}

After the hypersurface is partitioned into patches, we proceed to sampling particles from every patch independently. The sampling is already described in \cite{Oliinychenko:2019zfk}, but we repeat the description here for completeness.
We impose conservation laws in each patch, but allow variations of local  
energy density, quantum number densities, and collective velocities from cell to cell within a patch. These variations are characterized by the values of temperature, $T$, chemical potentials $\mu_B$, $\mu_S$, $\mu_Q$, and collective fluid velocity $u$ of the cells. For example, if a cell has a larger temperature or chemical potential it is more likely that a particle will be sampled from it. The local variations of collective velocity $u$ are important for a faithful description of higher order azimuthal anisotropies~\cite{Gardim:2011xv}, which otherwise would be smeared. This becomes obvious if one imagines a small system, such as $pp$ or $pPb$ collision, where the whole system may be one patch. The following multi-particle probability $P$ of a given particle configuration satisfies our requirements:
 \begin{eqnarray} \label{Eq:sampled_distribution}
P(N, \{N_s\}^{\mathrm{species}}, \{x_i\}_{i=1}^N, \{p_i\}_{i=1}^N) = \mathcal{N} \nonumber \\
\left( \prod_s \frac{1}{N_s!} \right) \prod_{i=1}^N \frac{g_i}{(2\pi\hbar)^3} \frac{d^3p_i}{p_i^0} p_i^{\mu}d\sigma_{\mu} \, \mathit{f}_i(p^{\nu}_i u_{\nu}, T, \mu_i) \times  \nonumber \\
\delta^{(4)}(\sum_i p^{\mu} - P^{\mu}_{tot}) \, \delta_{\sum_i B_i}^{B_{tot}}
\delta_{\sum_i S_i}^{S_{tot}} \, \delta_{\sum_i Q_i}^{Q_{tot}}
\end{eqnarray}
 It is a product of the usual Cooper-Frye formulas and global delta-functions which guarantee conservation laws over the patch. The $\frac{1}{N_s!}$ factors ensure that the Eq.~(\ref{Eq:sampled_distribution}) transforms into a standard microcanonical distribution, if our hypersurface is just one static cell. This property is crucial, because without it the sampling cannot be called microcanonical. Note that here the number of particles of each hadron species $N_s$ is not fixed, and neither is the total number of particles $N = \sum_s N_s$. Instead, both are distributed according to Eq.~(\ref{Eq:sampled_distribution}). The quantities $d\sigma_{\mu}$, $u^{\mu}$, $T$, and $\mu_{B,S,Q}$ depend on the spatial position of a particle $x_i$. The charges  $B_{tot}$, $S_{tot}$ and $Q_{tot}$ are computed using Eq.~(\ref{Eq:cons_laws}). It is important to underline, that the resulting sampled particles are defined by the distribution (\ref{Eq:sampled_distribution}), which should be the same regardless which algorithm is used to generate it.
 
 Sampling of the $N$-particle probability distribution expressed by Eq.~(\ref{Eq:sampled_distribution}) is generally difficult due to the unknown normalization factor $\mathcal{N}$ and the $\delta$-functions. We overcome this difficulty by applying a Metropolis algorithm, also known as a Markov chain Monte Carlo (MCMC) method, which in our case is closely related to solving the Boltzmann equation with the stochastic rate method~\cite{Seifert:2017oyb}. The state of our Markov chain $\xi$ depends on multiplicities, coordinates and momenta of all particles: $\xi = \xi(N, \{N_s\}^{\mathrm{species}}, \{x_i\}_{i=1}^N, \{p_i\}_{i=1}^N)$. The initial state is an arbitrary set of particles that satisfy the required conservation laws (Eq. \ref{Eq:cons_laws}). Quantum number conservation for the initial state is fulfilled by an \emph{ad hoc} heuristic algorithm picking the lightest particles, which can provide the required baryon number, strangeness, and electric charge. The energy-momentum conservation is achieved by rescaling the momenta as in \cite{Schwarz:2017bdg}. This initial state selection does not influence the resulting samples, because it is ''forgotten`` by Markov chain after a sufficient number of steps. Given a state $\xi$ we propose a state $\xi'$ with probability $T(\xi \to \xi')$ and then decide, if this state should be accepted, with probability $A(\xi \to \xi')$. Therefore, the probability to obtain a state $\xi'$ from $\xi$ is $w(\xi\to \xi') = T(\xi \to \xi') A(\xi \to \xi')$. The master equation, connecting the probability to obtain the state $\xi$ at steps $t$ and $t+1$ is

\begin{eqnarray}
P^{t+1}(\xi) - P^{t}(\xi) = \sum_{\xi'}  w(\xi' \to \xi) P^t(\xi') -\nonumber\\ w(\xi \to \xi') P^t(\xi) \,.
\end{eqnarray}

After many steps the probability $P^{t \to \infty}(\xi)$ should converge to $P(\xi)$ given by Eq. (\ref{Eq:sampled_distribution}). A sufficient condition for this is known as the detailed balance condition:

\begin{eqnarray}
\frac{P(\xi')}{P(\xi)} = \frac{w(\xi \to \xi')}{w(\xi' \to \xi)} =  \frac{T(\xi \to \xi')A(\xi \to \xi')}{T(\xi' \to \xi)A(\xi \to \xi')} \,.
\end{eqnarray}

This condition is satisfied if

\begin{eqnarray} \label{Eq:Metropolis_accept}
  a \equiv A(\xi \to \xi') = \textrm{min}\left(1, \, \frac{P(\xi') \, T(\xi' \to \xi)}{P(\xi) \, T(\xi \to \xi')}\right) \,.
\end{eqnarray}

 There is some freedom to select the proposal matrix $T(\xi\to\xi')$. We choose it such that it conserves energy, momentum, and quantum numbers. Consequently, our Markov chain never leaves the desired subspace where conservation laws are fulfilled. Our proposal matrix may be viewed as $2\to{3}$ and $3\to{2}$ stochastic ``collisions''~\cite{Seifert:2017oyb} on the hypersurface. However, we note, that there is no real time involved and ``collisions'' are not related to any physical process. They are simply a mathematical method  to sample the distribution of  Eq.~(\ref{Eq:sampled_distribution}). The proposal procedure is the following:
\begin{enumerate}
    \item With 50\% probability choose a $2\to 3$ or $3 \to 2$ transition.
    \item Select the ``incoming'' particles by uniformly picking one of all possible pairs or triples.
    \item Select the outgoing channel democratically with probability $1/N^{ch}$, where $N^{ch}$ is the number of possible channels, satisfying both quantum number and energy-momentum conservation.
    \item For the selected channel sample the ``collision'' kinematics uniformly from the available phase space $dR_{n}$ with probability $\frac{dR_n}{R_n}$, $n = 2$ or $3$.
    \item Choose a cell for each of the outgoing particles uniformly from all cells in the patch. Note that this choice affects the acceptance probability, because the corresponding temperatures, chemical potentials, velocities $u^{\mu}$, and normal 4-vectors $d\sigma_{\mu}$ in the Eq. (\ref{Eq:sampling_acceptance_probability}) will be taken at the cells, where the outgoing particles are thrown.
\end{enumerate}
Here $R_n$ is a phase-space integral for outgoing particles defined as the integral over $dR_n$:
\begin{eqnarray} \label{Eq:R_n_definition}
dR_n(\sqrt{s}, m_1, m_2, \dots, m_n) =  \frac{(2\pi)^4}{(2\pi)^{3n}}  \nonumber \\
\frac{d^3p_1}{2 E_1} \frac{d^3p_2}{2 E_2} \dots \frac{d^3p_n}{2 E_n} \delta^{(4)}(P_{tot}^{\mu} - \sum P_i^{\mu}) \,,
\end{eqnarray}
where $\sqrt{s} = (P_{tot}^{\mu} P^{tot}_{\mu})^{1/2}$. The integration of $dR_2$ and $dR_3$ is possible analytically \cite{Bauberger:1994nk,Seifert:2017oyb}. Our proposal procedure generates the following probabilities for $2\to 3$ and $3\to 2$ proposals:
\begin{eqnarray} \label{Eq:proposal_probabilities1}
T(2 \to 3) = \frac{1}{2} \frac{G_2^{ch}}{G_2} \frac{1}{N^{ch}_3} \frac{dR^{ch}_3}{R^{ch}_3} \frac{1}{N_{cells}^3}\\
\label{Eq:proposal_probabilities2}
T(3 \to 2) = \frac{1}{2} \frac{G_3^{ch}}{G_3} \frac{1}{N^{ch}_2} \frac{dR^{ch}_2}{R^{ch}_2} \frac{1}{N_{cells}^2}\,,
\end{eqnarray}
where $G_2 = \frac{N(N-1)}{2!}$ and $G_3 = \frac{N(N-1)(N-2)}{3!}$ denote total numbers of incoming pairs and triplets of any species, while $G_2^{ch}$ and $G_3^{ch}$ are the numbers of ways to select a given incoming particle species. Consequently, 
$\frac{G_2^{ch}}{G_2}$ and $\frac{G_3^{ch}}{G_3}$ represent the probabilities to obtain pairs and triplets of a given incoming species. The number of possible triplets and pairs of outgoing species with appropriate  quantum numbers are denoted by $N^{ch}_3$ and $N^{ch}_2$.
Inserting the proposal probabilities, Eqs.~(\ref{Eq:proposal_probabilities1}) and (\ref{Eq:proposal_probabilities2}), as well as the desired probability distribution, Eq.~(\ref{Eq:sampled_distribution}), into the expression for the acceptance probability,  Eq.~(\ref{Eq:Metropolis_accept}), we arrive, after some algebra, at  
\begin{eqnarray} \label{Eq:sampling_acceptance_probability}
a_{n\to m} = \frac{N^{ch}_m R_m}{N^{ch}_n R_n}  \frac{N!}{(N+m-n)!} \frac{m!}{n!} \frac{ k^{id}_m!}{ k^{id}_n!} \times \nonumber \\ \left( \frac{2 N_{cells}}{\hbar^3}\right)^{m-n} \frac{\displaystyle\prod_{i=1}^m g_i \, f_i(\mu_i - p_i^{\alpha}u_{\alpha},T) \, p^{\mu}_i d\sigma_{\mu}}{\displaystyle\prod_{j=1}^n g_j \, f_j(\mu_j - p^{\alpha}_j u_{\alpha},T) \, p^{\mu}_j d\sigma_{\mu}}
\end{eqnarray}
where we made use of the relation  $\prod \frac{d^3p_i}{(2\pi\hbar)^3 p^0_i} \delta^{(4)}(P_{tot}^{\mu} - \sum P_i^{\mu}) = 2^n \frac{dR_n}{(2\pi)^4} $. Here $n=2,3$ and $m=3,2$ are the numbers of incoming and outgoing particles, and $N$ is the total number of particles before proposing the Markov chain step. The product in the numerator is taken over the outgoing particles and the one in the denominator is taken over the incoming particles. The quantities $d\sigma$, $u$, $T$, $\mu$ should be evaluated in the cell where the particles are proposed to be, or coming from. The  total number of particles in the entire patch is given by $N$, and $k^{id}_m$ and $k^{id}_n$ are the numbers of outgoing and incoming identical species in the reaction. Note that the sampling accounts for  the variations in temperature and chemical potential within the patch. Also, and equally important, the distribution function $f$ may contain viscous corrections. To summarize, the algorithm consists of multiple Markov chain steps, where the step is proposed according to Eqs.~(\ref{Eq:proposal_probabilities1}) and (\ref{Eq:proposal_probabilities2}) and accepted with probability given by Eq.~(\ref{Eq:sampling_acceptance_probability}).

Testing and validation of the sampling is performed in the appendices \ref{appendix:massless_microcanonical} and \ref{appendix:Begun_microcanonical}, as well as in Ref. \cite{Oliinychenko:2019zfk}.

\subsection{Convergence and runtime} \label{sec:runtime}

Our goal is to generate $N_{samples}$ samples from the distribution (\ref{Eq:sampled_distribution}) as fast as possible, but in such a way that they are not correlated with each other. In addition these samples should not depend on the \emph{ad hoc} initial state of the Markov chain. The last two requirements imply a sufficient (and the larger the better) number of Markov chain steps. The runtime minimization, however, demands the minimal number of Markov chain steps. Here we describe our approach to address this problem, which focuses more on robustness rather than runtime minimization.

After the generation of the initial state of our Markov chain, we perform a warm-up of $N_{warmup}$ steps described above to reach equilibration. Because the warm-up is performed only once per one hydrodynamic hypersurface, we simply play it safe, set a large $N_{warmup} = 10^6$, and check that it provides distributions, that do not change if one increases $N_{warmup}$. Then the resulting particles are printed out.

The next sample should not be correlated with the previous one. This is achieved by performing $N_{decor}$ steps between printing out the sample. After this it is not clear if the required decorrelation is reached. Insufficient decorrelation mainly exhibits itself as spikes in momentum spectra, which typically occur at the high momentum tail of the distribution. These spikes originate from one or two particles ``stuck'' in a corner of momentum space for many Markov chain steps. To get rid of these spikes, we perform additional $N_{decor}$ ``$2 \leftrightarrow 2$ elastic'' steps described further. Then we check if there are any particles unchanged after these steps. In case there are unchanged particles we perform $N_{decor}$ $2 \leftrightarrow 2$ elastic steps again and repeat these blocks of $N_{decor}$ $2 \leftrightarrow 2$ elastic steps until all particles are changed. Then we print out the resulting particles. The whole procedure is repeated as many times as many events we need. In our calculations we used $N_{decor} = 200$ and $N_{decor} = 500$ and did not observe any difference in results. 
For tests presented in the  Appendices and tests in \cite{Oliinychenko:2019zfk} $N_{decor} = 500$ was used.

The  $2 \leftrightarrow 2$ elastic steps are the Markov chain steps, where particles of the same species are proposed, but their cells and momenta are allowed to change. They are used for decorrelation for a single reason: they cannot bias multiplicity distributions, because they do not change any multiplicities. In contrast, repeating $2 \leftrightarrow 3$ decorrelation blocks \emph{until all particles are changed} may bias multiplicity distributions. As an extreme example, consider a patch with only 2 particles, and set $N_{decor} = 1$. Forcing $2 \leftrightarrow 3$ decorrelation steps until all particles are changed means that the next sample always contains 3 particles even though the mean number of particles can be set arbitrarily close to 2. To avoid this type of bias we adopt $2 \leftrightarrow 2$ elastic steps for decorrelation. The acceptance probability of $2 \leftrightarrow 2$ elastic steps is expressed by Eq. (\ref{Eq:Metropolis_accept}), with $m = n = 2$ and most of the factors cancelling, resulting in

\begin{eqnarray} \label{Eq:sampling_acceptance_probability22}
a_{2\to 2} =  \frac{\displaystyle\prod_{i=1}^2 f_i(\mu_i - p_i^{\alpha}u_{\alpha},T) \, p^{\mu}_i d\sigma_{\mu}}{\displaystyle\prod_{j=1}^2 f_j(\mu_j - p^{\alpha}_j u_{\alpha},T) \, p^{\mu}_j d\sigma_{\mu}} \,,
\end{eqnarray}

where the product in the numerator is over the outgoing particles, and product in denominator is over the incoming ones. As in Eq. (\ref{Eq:sampling_acceptance_probability}) the quantities $d\sigma$, $u$, $T$, $\mu$ should be evaluated in the cell where the particles are proposed to be, or coming from.

Mainly due to the decorrelations our sampling procedure appears to be rather time-consuming. The dependencies on the runtime on the parameters of the problem are rather peculiar. It does not depend on the number of cells in the patch, unlike for the usual grand-canonical sampler. Neither it depends on the number of the sampled species. It depends indirectly on the acceptance rate, because if the acceptance rate is low, the decorrelation will take more steps. For realistic hypersurfaces we have observed acceptance rate of $5-10$ \%, and it can decrease, if the patches are very non-uniform.

For our realistic hypersurface from Au+Au collisions at 19.6 GeV on a single Intel Xeon 2.4 GHz processor we found the runtime scaling as
\begin{eqnarray}
 t \mathrm{ [s]} = N_{patches} \frac{N_{samples}}{10^4} \left( 48 + \left(\frac{E_{patch }\mathrm{ [GeV]}}{7.8}\right)^2\right) \,,
\end{eqnarray}
or in other words, it takes about one minute per patch per $10^4$ events for $E_{patch} = 20$ GeV. As the sampling in every patch is performed independently, we have parallelized our code over the patches. The quadratic dependency on $E_{patch}$ is due to decorrelation, therefore a way to speed up the sampling dramatically is to relax the decorrelation requirements, which in our case are very strict.  Another possible idea is to consider $N \to N$ elastic steps for decorrelation instead of $2 \to 2$.

\subsection{Negative contributions} \label{sec:negative_contributions}

\begin{figure*}
    \centering
    \includegraphics[width=0.9\textwidth]{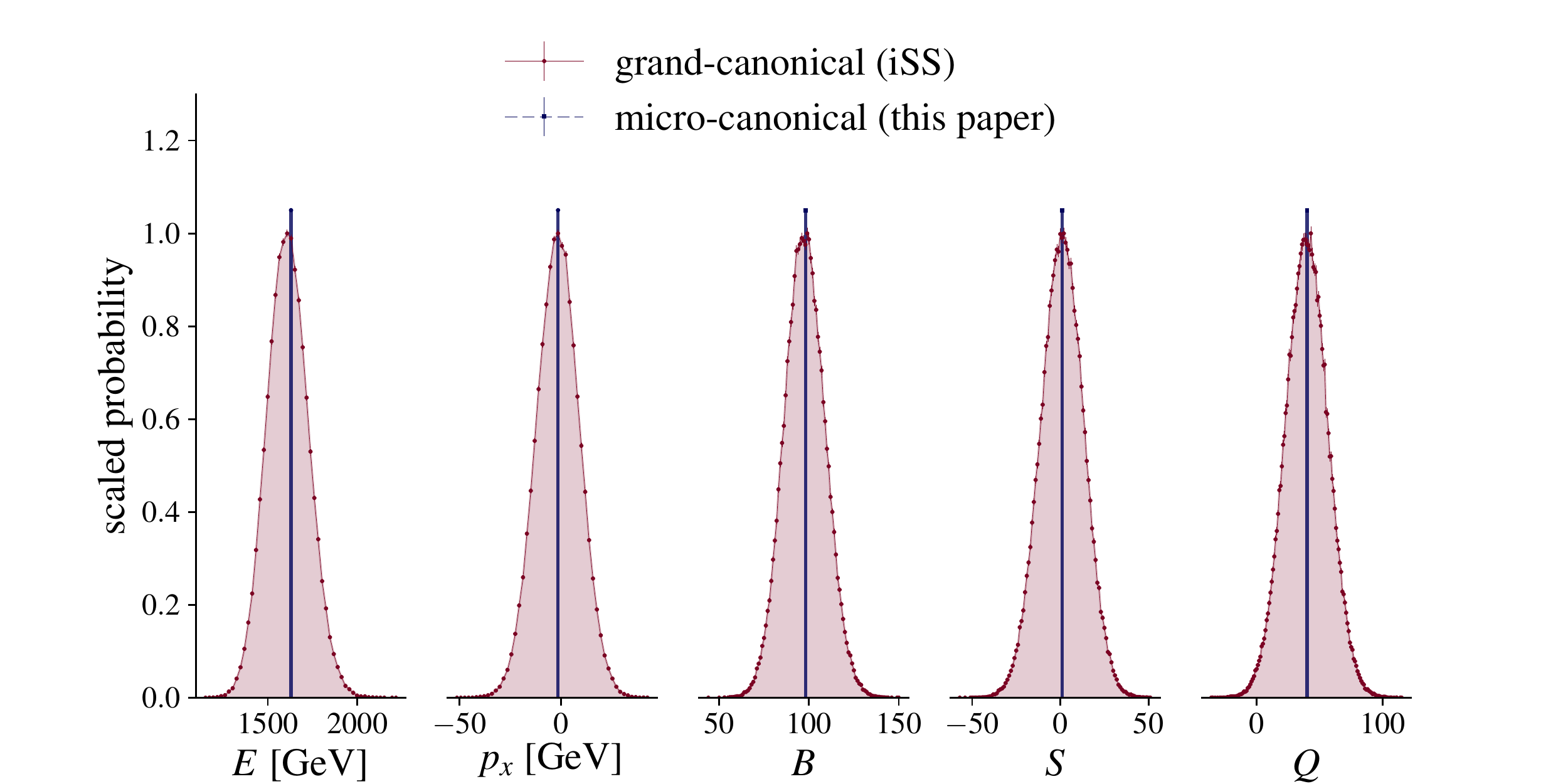}
    \caption{Distributions of total energy, x-component of momentum, net baryon number, net strangeness, and net charge of all sampled particles. Our microcanonical sampler is compared to grand-canonical one. By construction, quantities from microcanonical sampler are identical to those of the hydrodynamic hypersurface in each event, while the quantities from the grand-canonical sampler are distributed around them. Probability distributions are scaled for viewing convenience to have maxima of 1 (grand-canonical) and 1.05 (microcanonical).}
    \label{fig:conserved_total_distribution}
\end{figure*}

Our microcanonical sampler currently treats the long-standing problem  of negative Cooper-Frye contributions \cite{Bugaev:1996zq,Bugaev:1999wz,Grassi:2004dz} in a special way, different from a typical grand-canonical sampler. To set the stage, let us first explain the problem. Grand-canonical samplers use the Cooper-Frye formula to compute how many particles should be produced from a cell with with a four-volume $d\sigma_{\mu}$ at given momentum:
\begin{eqnarray} \label{eq:dn_cooper_frye}
dN \sim \frac{g_i}{(2\pi\hbar)^3} \frac{d^3p_i}{p_i^0} p_i^{\mu}d\sigma_{\mu} \, \mathit{f}_i(p^{\nu}_i u_{\nu}, T, \mu_i)
\end{eqnarray}
The factor $p_i^{\mu}d\sigma_{\mu}$ is negative for particles that cross the hypersurface inwards. These negative contributions are necessary to conserve energy, momentum, and charges across the hypersurface. However, they are not possible to sample, because they come with negative weights. Moreover, integrating Eq.~(\ref{eq:dn_cooper_frye}) over momenta (as it is done for example in the Appendix \ref{appendix:integration}) one can see that the net particle flow is proportional to $u^{\mu} d\sigma_{\mu}$, which can also be negative.  Therefore, the usual solution is:
(1) ignore cells with negative $u^{\mu} d\sigma_{\mu}$, which means that the net particle flow is directed inwards. In this way both positive and negative contributions from these cells are neglected; and (2) for cells with positive $u^{\mu} d\sigma_{\mu}$ sample only particles with positive $p_i^{\mu}d\sigma_{\mu}$, which formally corresponds to multiplying distribution by $\theta(p_i^{\mu}d\sigma_{\mu})$. This cuts off negative energy flow, or energy flow inwards the hypersurface. To summarize, the usual Cooper-Frye sampler ignores the inward flow of energy, as well as the outward flow originating from cells with negative $u^{\mu} d\sigma_{\mu}$. The same is valid for the flow of momenta and charges ($B$, $S$, $Q$). As a consequence, conservation laws are violated even on average by events, unless the sampler is intentionally modified to avoid this, such as in \cite{Huovinen:2012is}. However, existing modifications of this kind are ad hoc \cite{Schwarz:2017bdg} and do not reproduce a canonical or microcanonical ensemble in a box.

We have encountered a practical example of the negative contributions problem, when we tried to use a hypersurface from MUSIC hydrodynamics with dynamical initialization \cite{Shen:2017bsr} at 19.6 GeV and 30-40\% centrality. This initial state results in a highly irregular particlization hypersurface on which the ratio of the total particle flow inwards over the net particle flow, $\sum u^{\mu}d\sigma_{\mu} \theta(-u^{\mu}d\sigma_{\mu}) / \sum u^{\mu}d\sigma_{\mu}$, constitutes around -23\%. For this case we obtain around 20\% smaller energy from particles generated by iSS  than that of the hypersurface. 

Our microcanonical sampler deals with the negative contributions in the following way. The conserved quantities are computed first according to the Eq. (\ref{Eq:cons_laws}), where negative contributions are present. Therefore, the conserved quantities of the hypersurface are equal to the ones of the sampled particles, unlike in the grand-canonical sampler. Then, the sampled particles according to the Eq. (\ref{Eq:sampled_distribution}) are allowed to fly inwards. In other words, the negative contributions are actually sampled. This is possible, because the $p^{\mu} d\sigma_{\mu}$ factors, which can be negative, enter the multi-particle probability distribution (Eq. \ref{Eq:sampled_distribution}) as a product. While the product should be positive, the sign of individual $p^{\mu} d\sigma_{\mu}$ is not restricted, and we are thus able to sample negative contributions.

\begin{figure*}
    \centering
    \includegraphics[width=0.2445\textwidth]{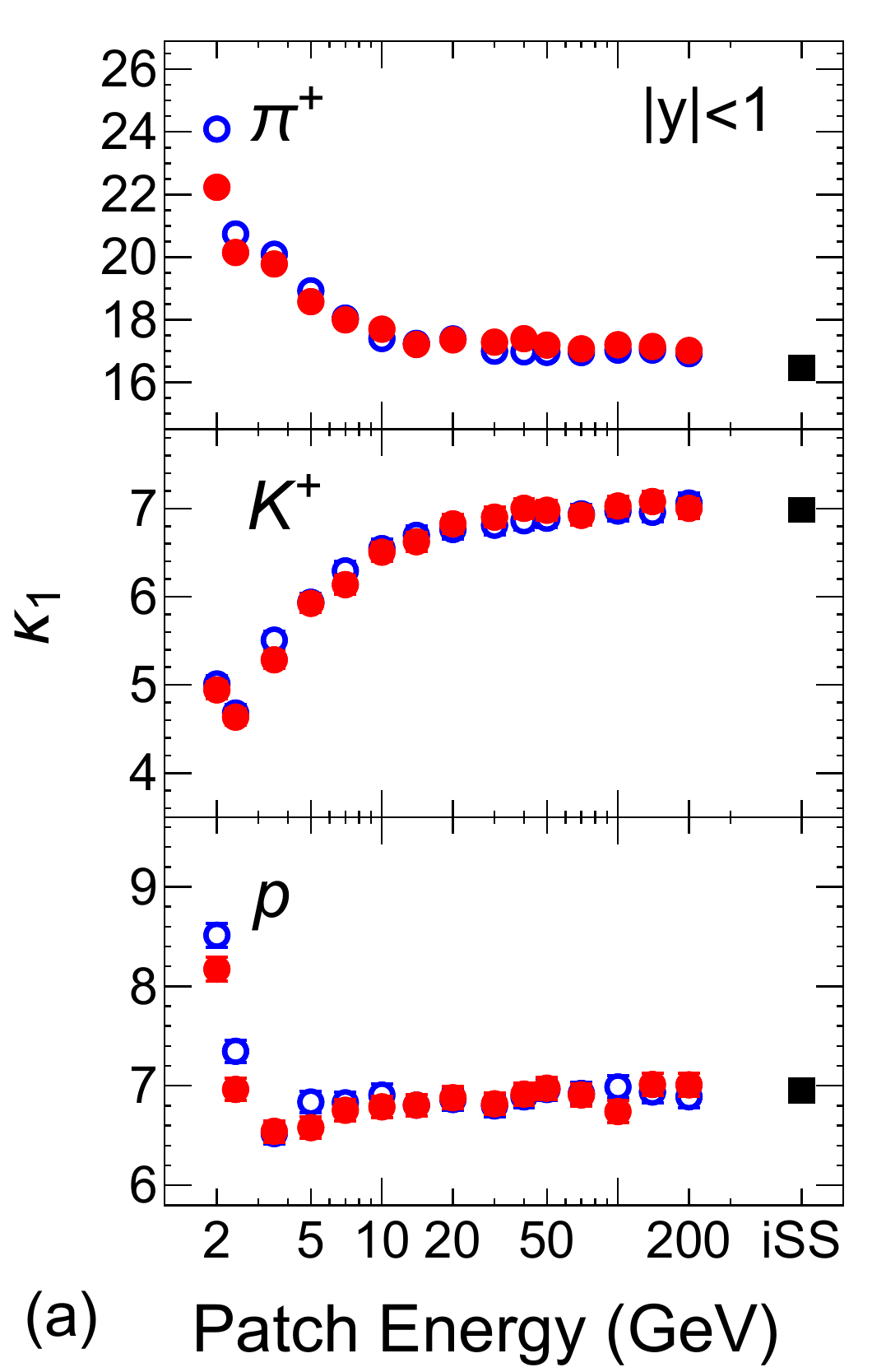}
    \includegraphics[width=0.2445\textwidth]{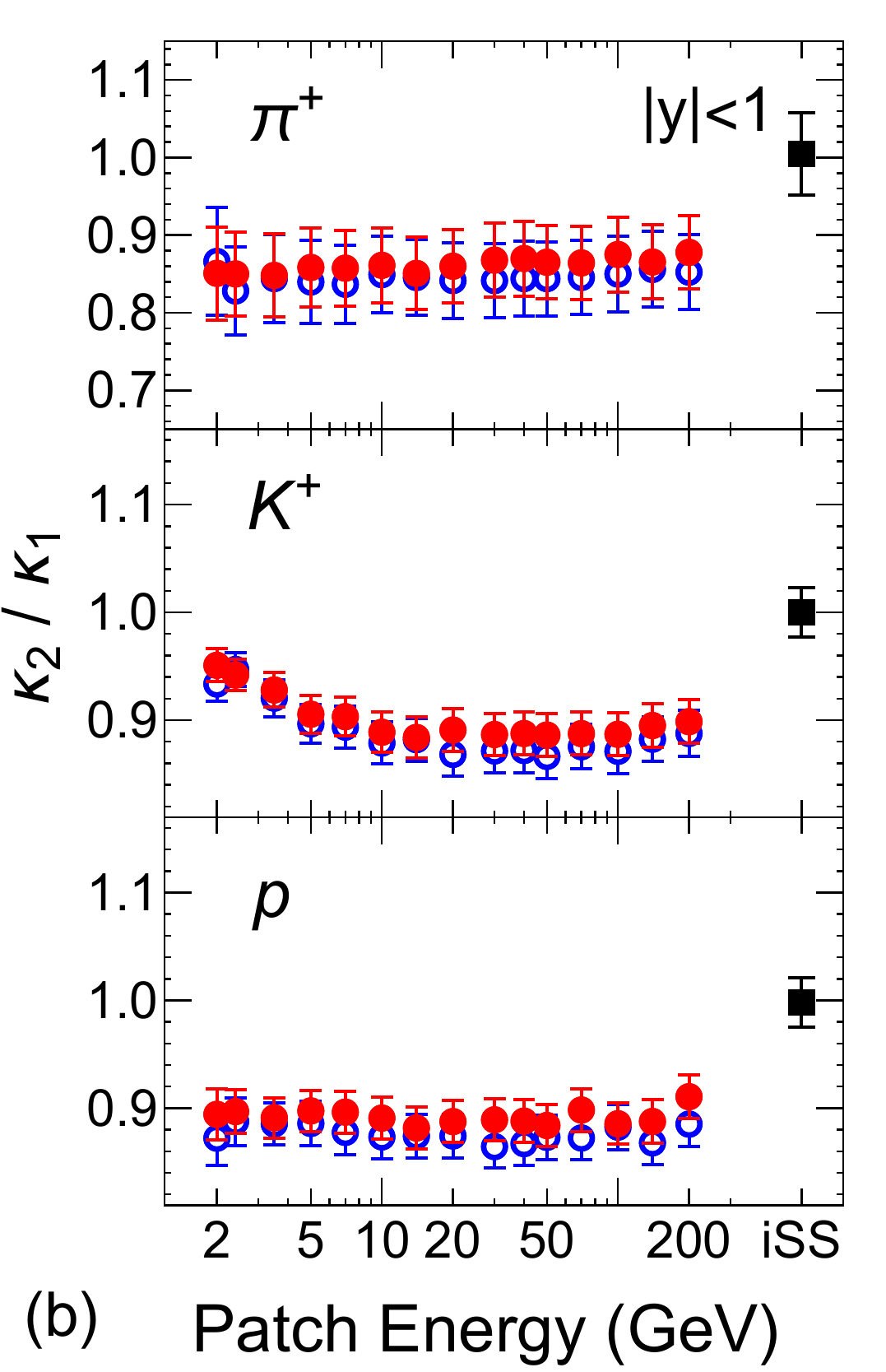}
    \includegraphics[width=0.2445\textwidth]{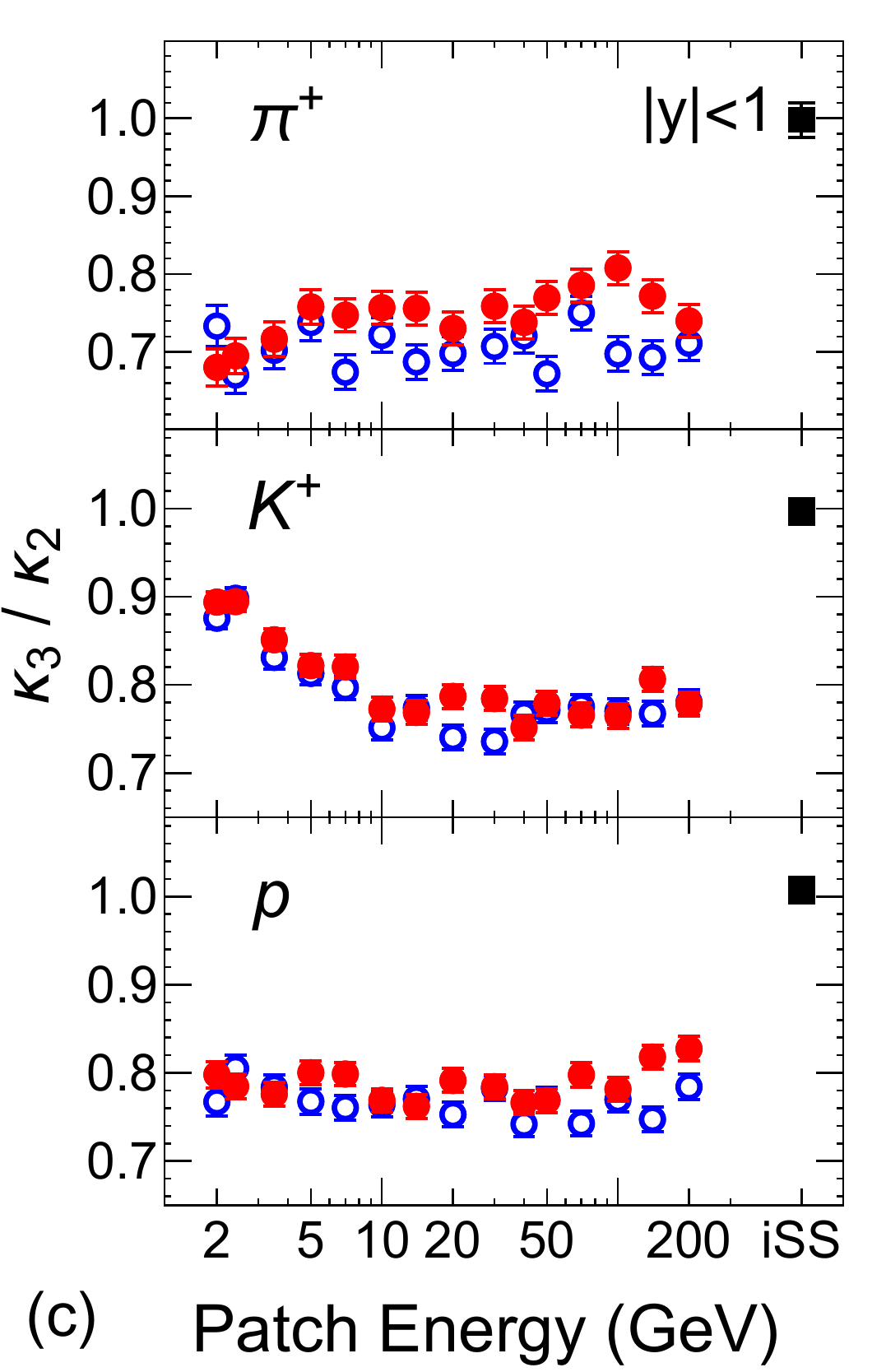}
    \includegraphics[width=0.2445\textwidth]{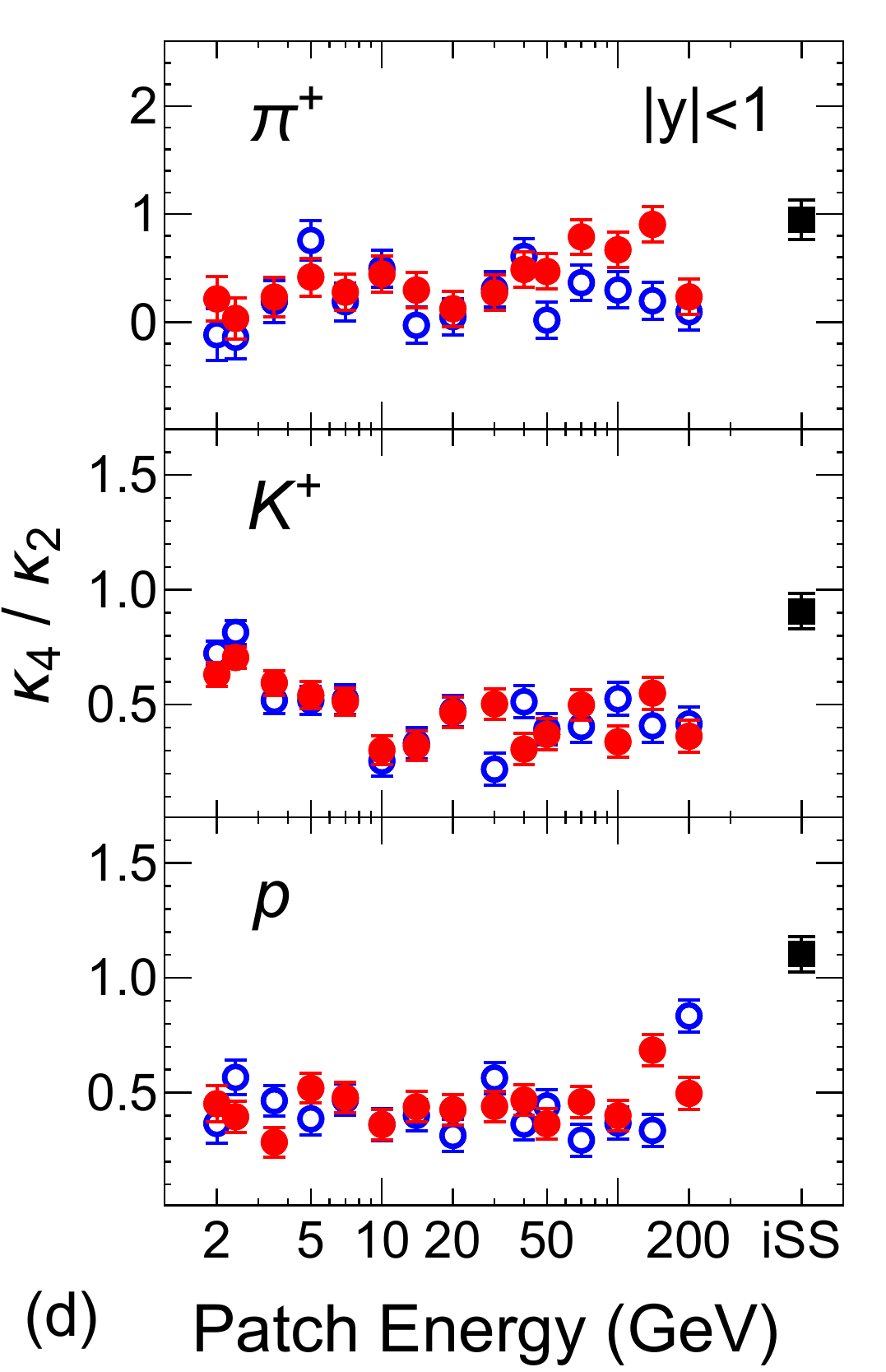}
    \caption{Mean multiplicities (a), scaled variances (b), and ratios of cumulants $\kappa_3/\kappa_2$ (c) and $\kappa_4/\kappa_2$ (d) of $\pi^+$, $K^+$, and $p$ multiplicities within rapidity range $|y|<1$. Our microcanonical sampler (red circles) as a function of the patch energy is compared to the iSS grand-canonical sampler (black squares). Closed and open symbols are results for different patch splitting algorithms: closed --- largest energy cell and $\Delta r^2/d_0^2 + (\Delta T / \sigma_T)^2 + (\Delta \mu_B / \sigma_{\mu_B})^2$ distance, open --- largest $\eta$ cell and $\Delta \eta$ distance.}
    \label{fig:mean_scaledvar_particle}
\end{figure*}

At first glance it seems to be the solution of a long-standing negative contributions problem, but unfortunately it is not. This approach is suitable for particlization of pure hydrodynamics, but not necessarily for a hybrid simulation. One can see this immediately, if one considers a Sun-like object, as proposed in \cite{Bugaev:1999wz}. The hypersurface there is a static sphere with $u^{\mu} = (1, 0,0,0)$, therefore, the number of particles crossing it inwards and outwards are equal. Then our microcanonical sampler correctly computes that the net energy flow across the hypersurface is 0, and does not sample any particles. In a  simulation we need a different outcome: sampling particles going outwards, and absorbing particles from transport going inwards as source terms. This cannot be achieved by sampling alone. It should be done by matching hydrodynamics with transport and solving them together. The sampling in this case should include $\theta$-functions in Eq. (\ref{Eq:cons_laws}) and in the probability in Eq. (\ref{Eq:sampled_distribution}). For this reason we leave the further investigation of negative contributions for the subsequent work. In this work we use a hypersurface, for which negative contributions are negligible. This allows a fair comparison between grand- and microcanonical samplers.

\section{Application to a realistic hypersurface} \label{sec:results}

In this section, we apply the microcanonical sampler for its main intended use case --- particlization of hydrodynamics in heavy ion collisions simulations. This serves three purposes: first, it is a comprehensive test of both the sampler performance and patch splitting procedure; second, it allows to demonstrate the sensitivity of observables to the patch size; third, it allows to study correlations due to conservation laws as a function of kinematic cuts, such as measured recently in \cite{Adam:2019xmk}.

For these purposes we consider a typical particlization hyper-surface from $30-40\%$ mid-central Au+Au collisions at $\sqrt{s_{NN}}=19.6$~GeV. It is computed by a $3+1$~D MUSIC hydrodynamic simulation~\cite{Schenke:2010nt} with event-averaged Monte--Carlo Glauber initial condition. The hyper-surface corresponds to constant energy density of 0.4 GeV/fm$^3$. This is the same setup as in Ref.~\cite{Denicol:2018wdp}. The idea behind choosing $\sqrt{s_{NN}}=19.6$~GeV is to have a hypersurface large enough to demonstrate the capabilities of the sampler, but small enough to be able to generate the statistics necessary for computing higher order fluctuations. Partitioning of the hypersurface into patches was performed 10 times, and for each such partitioning $2 \cdot 10^4$ samples were generated, therefore the total number of samples is $N_{ev} = 2 \cdot 10^5$. Smoothed event-averaged initial condition is particularly important, because it leads to a smooth particlization surface with negligible negative Cooper-Frye contributions. This allows for a fair comparison between micro- and grand-canonical samplers, which  treat negative contributions in different ways, see Sec. \ref{sec:negative_contributions}.

First of all we demonstrate that the conservation laws over the hypersurface are indeed fulfilled in our sampler. In Fig. \ref{fig:conserved_total_distribution} we compare the distribution of total energy, x-component of momentum, net baryon number, net strangeness, and net electric charge from our microcanonical sampler and from the grand-canonical sampler \texttt{iSS} described and tested in \cite{Shen:2014vra} and available publicly at \cite{url:iSS}. The sampled particle species are identical for both samplers. Our microcanonical sampler is currently not able to produce quantum distributions, therefore, for a fair comparison we adjusted the standard iSS sampler to produce Boltzmann distribution instead of the default Bose and Fermi distributions. One can see in Fig. \ref{fig:conserved_total_distribution} that the average values of conserved quantities coincide, but for the microcanonical sampler quantities do not fluctuate event-by-event. This is the distinguishing feature of the microcanonical sampler, which follows by construction from Eq. (\ref{Eq:sampled_distribution}). The coincidence of the means is, however, not perfect: in Fig. \ref{fig:conserved_total_distribution} one can notice a small mismatch between mean energies, of the order of  1.5\%. This effect is statistically significant and originates from negative contributions, which are treated differently in the iSS and in our sampler (see Sec. \ref{sec:negative_contributions}). For a hypersurface with larger negative contributions this discrepancy becomes larger.

\begin{figure*}
    \centering
    \includegraphics[width=0.45\textwidth]{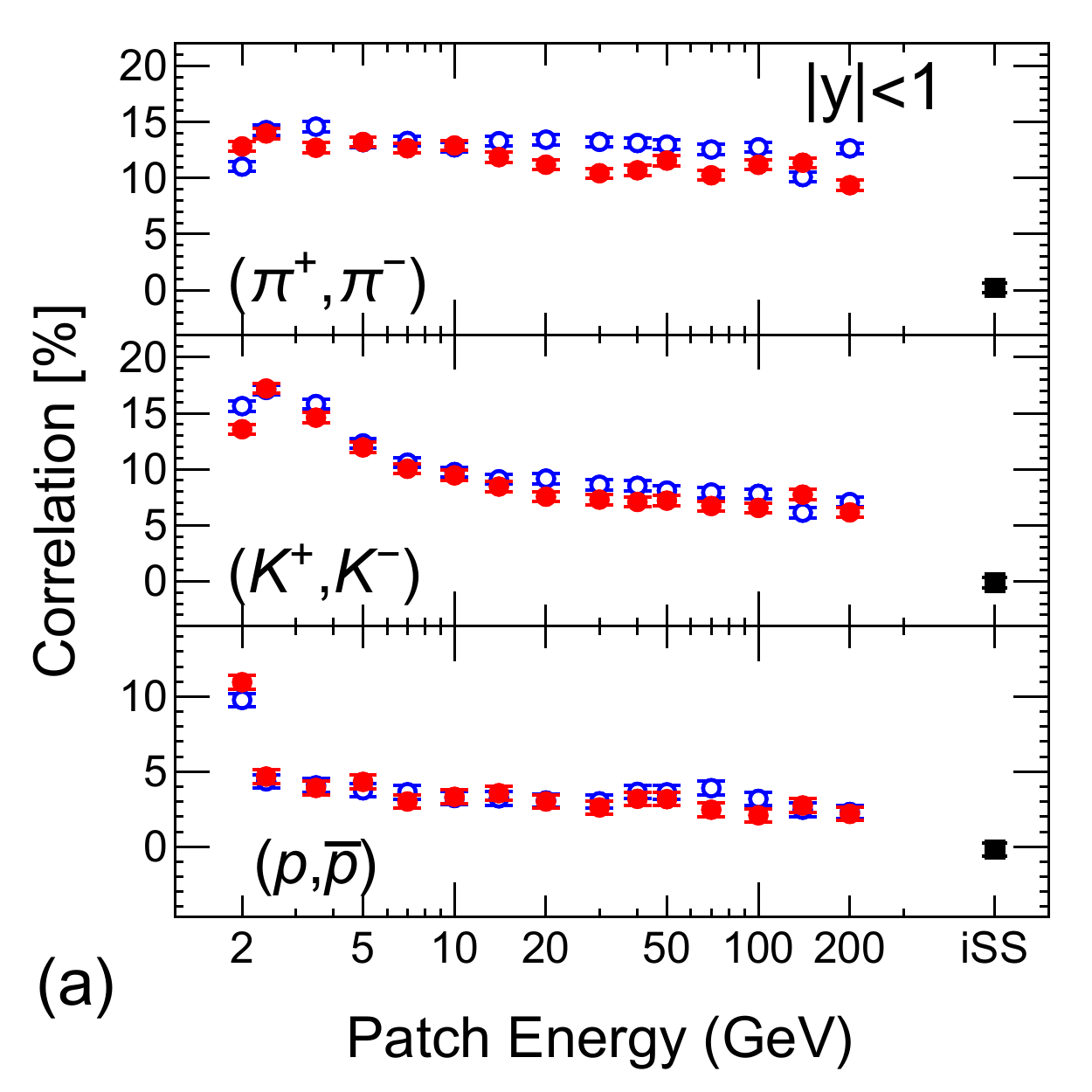}
    \includegraphics[width=0.45\textwidth]{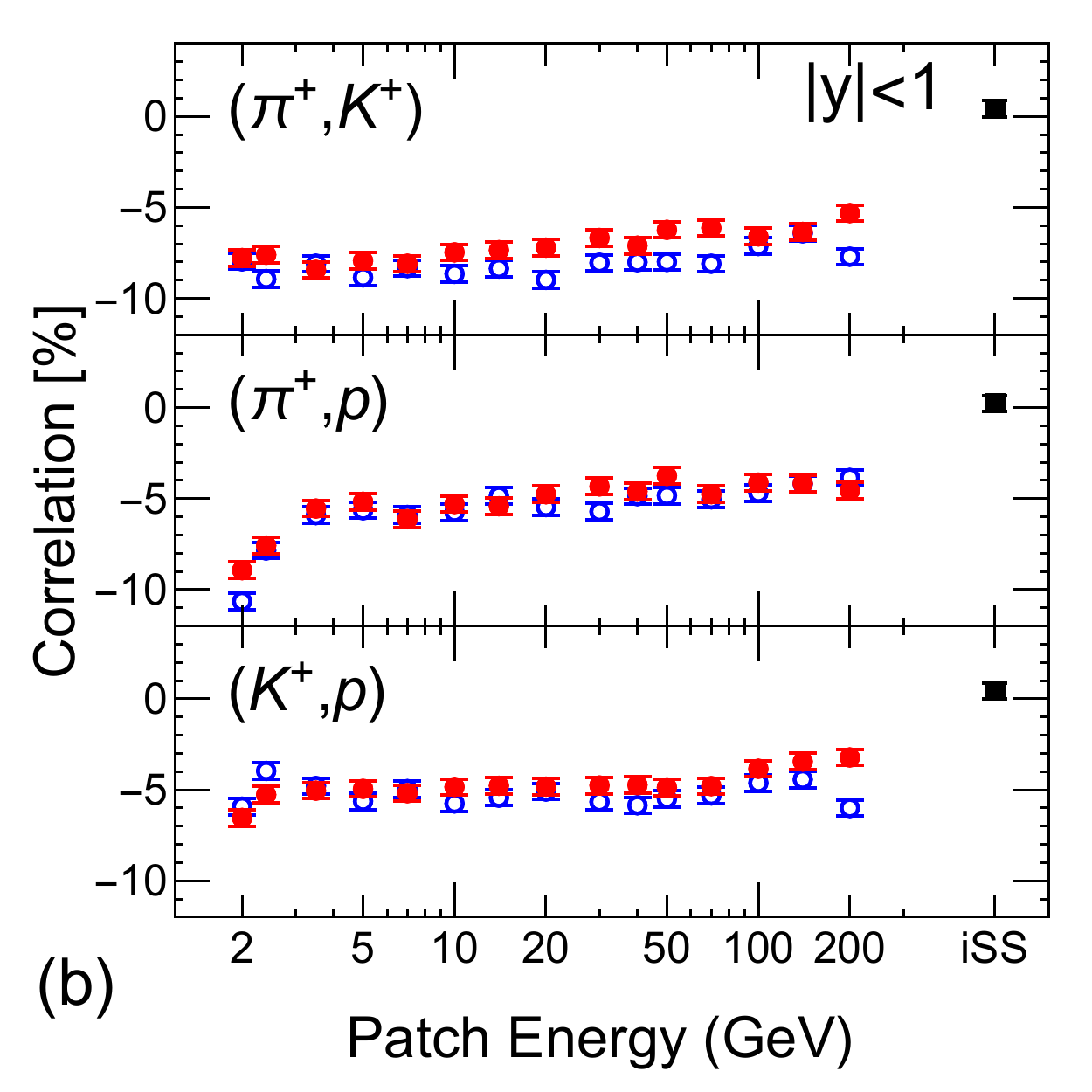}
    \caption{Correlations between particle multiplicities at midrapidity ($|y|<1$) as a function of patch energy. Our microcanonical sampler (red circles) is compared to the iSS grand-canonical sampler (black squares). Closed and open symbols are results for different patch splitting algorithms: closed --- largest energy cell and $\Delta r^2/d_0^2 + (\Delta T / \sigma_T)^2 + (\Delta \mu_B / \sigma_{\mu_B})^2$ distance, open --- largest $\eta$ cell and $\Delta \eta$ distance.}
    \label{fig:corr}
\end{figure*}

\subsection{Means, variances, correlations and fluctuations of $\pi$, $K$, $p$ at mid-rapidity}

The effects of the local microcanonical sampler are most evident as one varies the patch energy. For a single patch, these effects can be in principle computed analytically~\cite{Begun:2004pk}. However, if one studies particle distribution with a kinematic cut or acceptance window, then particles originate from many patches. In this case it is not clear a priori how much of these effects are preserved. Indeed, if one chooses a small subsystem of a microcanonical system, the subsystem will be grand-canonical. Therefore, a particular question that we want to address here is to which extent microcanonical effects are preserved if a kinematic cut is imposed. For this purpose we impose a $|y| < 1$ rapidity cut and consider different multiplicity cumulants up to fourth order $\kappa_{1-4}$. These cumulants and  their ratios are convenient to characterize multiplicity distributions: $\kappa_1 = \bar{N}$ is the mean of the distribution and  $\kappa_2/\kappa_1 = \sigma^2/\bar{N}$ represents the scaled variance. For the Poisson distribution, which is the multiplicity distribution associated with a grand-canonical sampling, all cumulants are equal to the mean particle number, and, therefore, the cumulant ratios are $\kappa_3/\kappa_2 = \kappa_4/\kappa_2 = 1$. Deviations of the cumulant ratios from unity demonstrate the magnitude of the microcanonical effects. It is important to note here, that we do not consider resonance decays in this work to isolate the effects of conservation laws.

In Fig.~\ref{fig:mean_scaledvar_particle} we show the mean value  (panel (a)) and standard deviation (panel (b)) of the multiplicity distributions for identified particles within the rapidity range $|y|<1$. A clear trend can be observed: with decreasing patch energy the number of pions increases while those of kaons and protons decrease. The reason is the following: when the particles are created in sub-volumes with less energy, the lighter ones are more favored. With increasing patch energy the averages approach the iSS grand-canonical values. However, for pions even for the largest patch energy there is a small difference between the microcanonical and grand-canonical means. This difference originates in the same way, as in Figs. \ref{fig:A1} and \ref{fig:B1}: even in the thermodynamic limit microcanonical means tend to be larger than the grand-canonical means, even though their ratio approaches unity. Scaled variances in the microcanonical case are systematically smaller than for iSS. The effect is almost independent on the patch size, constitutes around 10\%, and originates mainly from conservation of quantum numbers. This is a known (micro)canonical suppression of fluctuations. Similar result is obtained analytically in the thermodynamic limit \cite{Begun:2004gs}. In panels (c) and (d) one can see a less studied effect: microcanonical suppression of the higher order fluctuations. In the grand-canonical case the scaled skewness $\kappa_3/\kappa_2$ and kurtosis $\kappa_4/\kappa_2$ are always unity, but in the microcanonical case they turn out to be always below unity. Similarly to the second order fluctuations, the effect does not vanish even in the thermodynamic limit.

Next we consider correlations between various particles, where the correlation between quantities $A$ and $B$ in Fig. \ref{fig:corr} is defined as
\begin{equation}
    \mathrm{Corr}(A,B) \equiv \frac{\langle(A-\overline{A})(B-\overline{B})\rangle}{\sigma_A \,\sigma_B}
  \end{equation}
  In a grand-canonical sampler like \texttt{iSS}, particles are sampled independently, hence the multiplicity correlations always vanish. However, the micro-canonical sampler introduces non-vanishing correlations due to conservation laws. This is shown in Fig. \ref{fig:corr}. If conservation laws are more local (and the patch energy is smaller) correlations are larger. However, correlations do not vanish even in the thermodynamic limit. Although the correlation between multiplicities with a rapidity cut is less strong than without a cut, they remain to be significant, typically from 5 to 10\% in absolute value.
  The sign of the correlations in Fig. \ref{fig:corr} is evident already from the pure electric charge conservation, although baryon number and strangeness influence the magnitude significantly.

\begin{figure*}
    \centering
    \includegraphics[width=0.2445\textwidth]{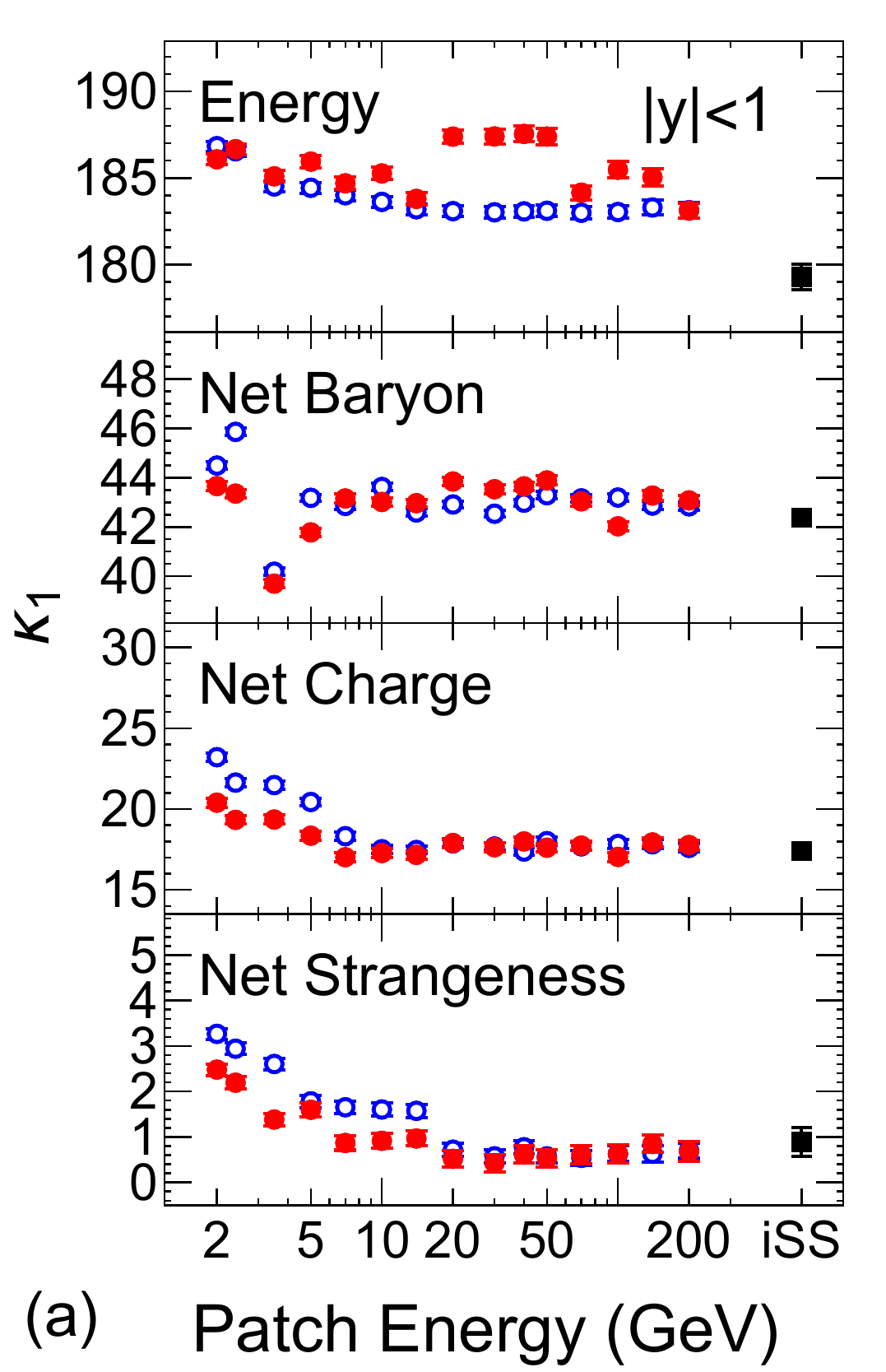}
    \includegraphics[width=0.2445\textwidth]{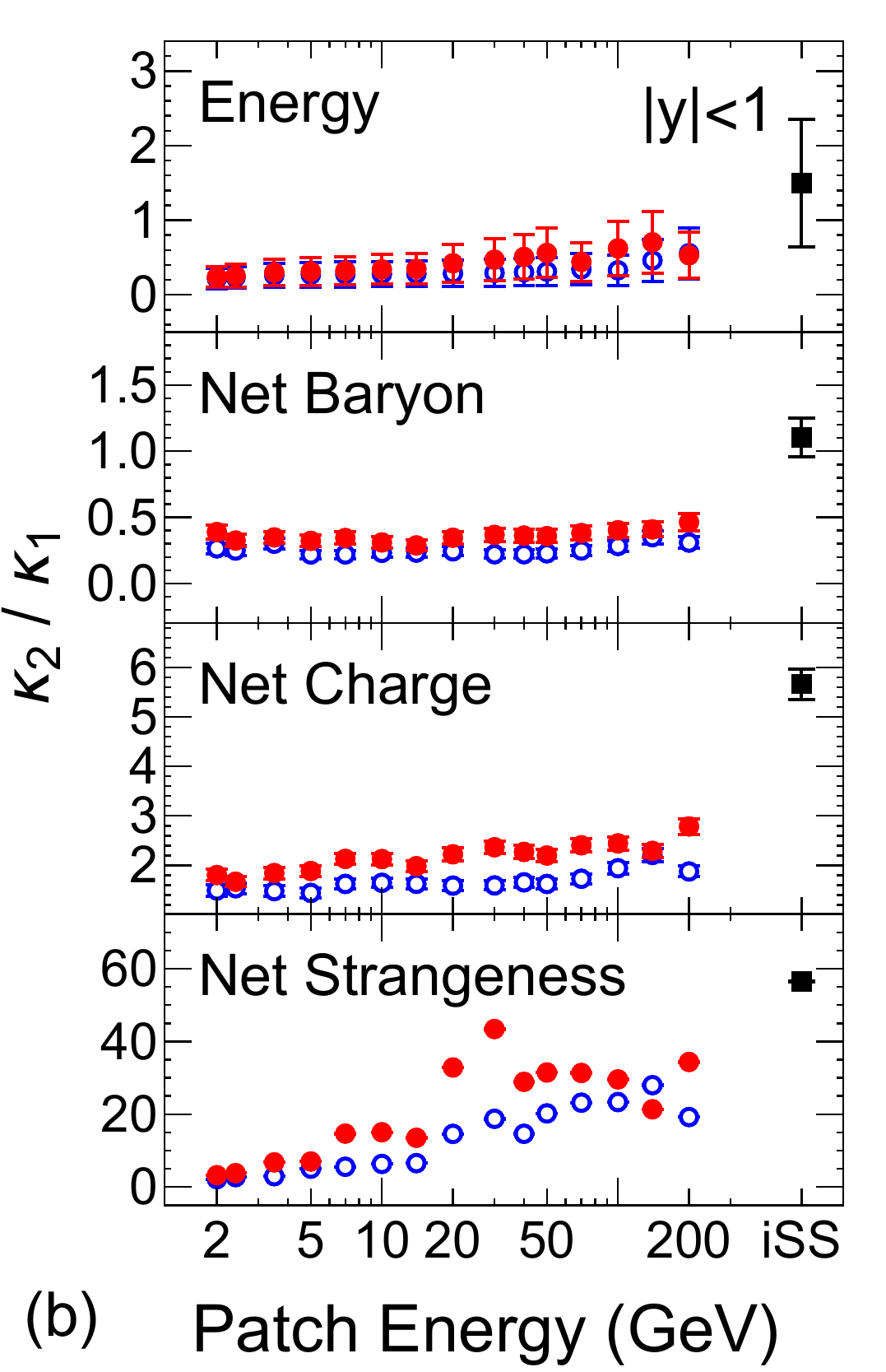}
    \includegraphics[width=0.2445\textwidth]{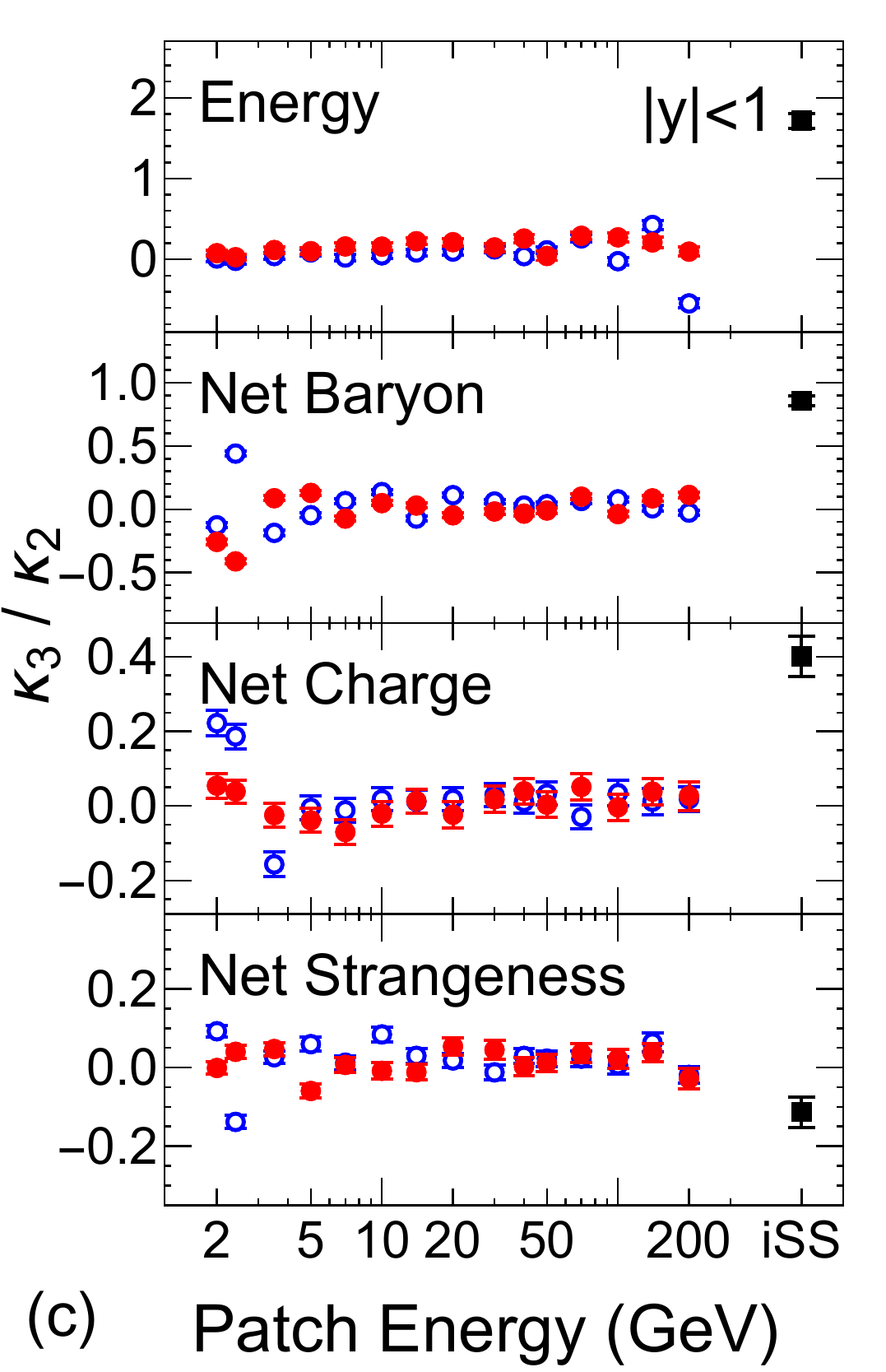}
    \includegraphics[width=0.2445\textwidth]{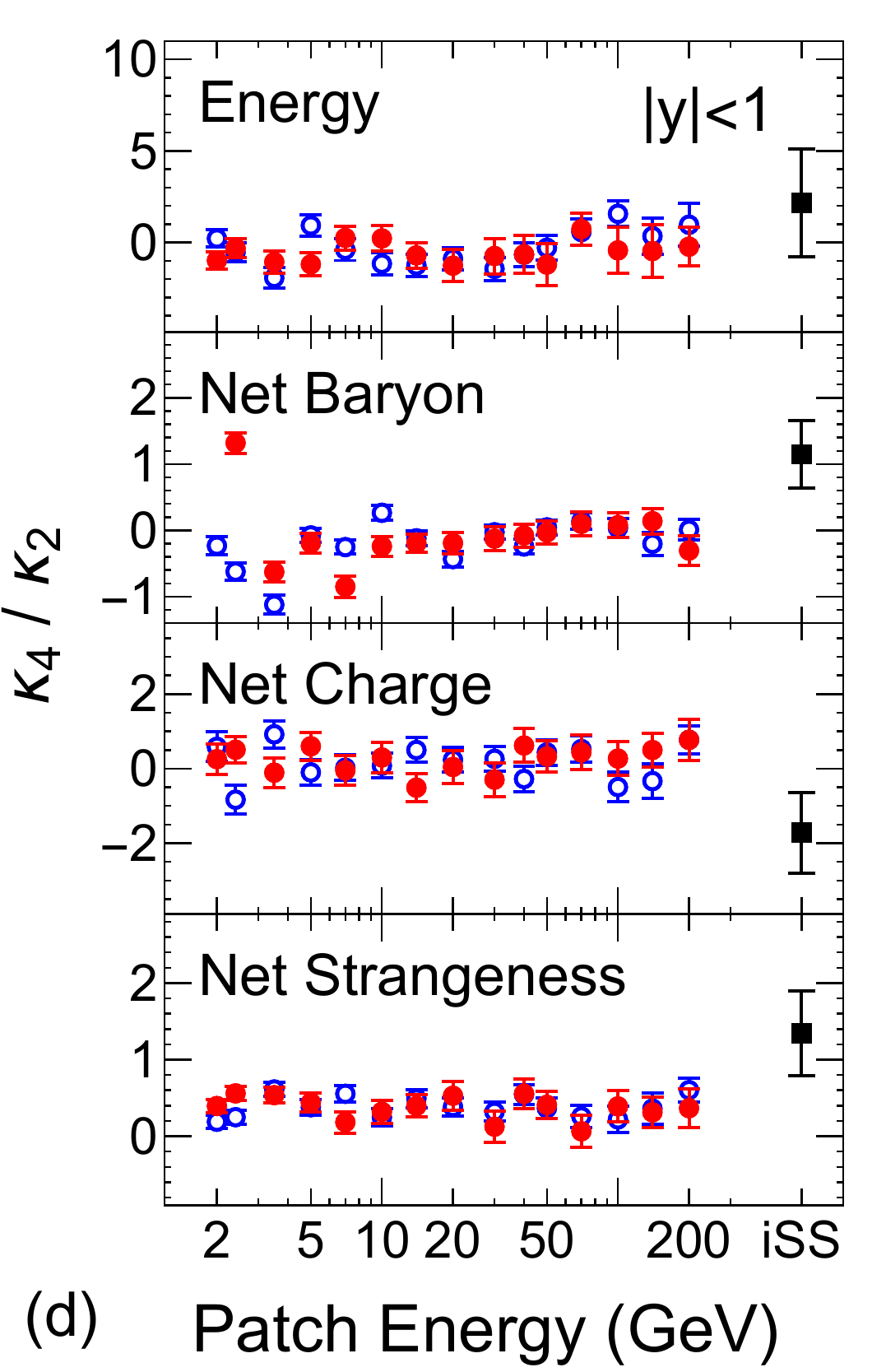}
    \caption{Cumulants or cumulant ratios of conserved quantities. From top to bottom: energy, net baryon number, net electric charge, and net strangeness, respectively. Particles are selected within a rapidity range $|y|<1$. Closed and open symbols are results for different patch splitting algorithms: closed --- largest energy cell and $\Delta r^2/d_0^2 + (\Delta T / \sigma_T)^2 + (\Delta \mu_B / \sigma_{\mu_B})^2$ distance, open --- largest $\eta$ cell and $\Delta \eta$ distance.}
    \label{fig:conserved_mean_and_fluctuations}
\end{figure*}

\subsection{Cumulants of conserved quantities within a rapidity cut}

We next study the fluctuations of conserved quantities, i.e. energy, net baryon number, net electric charge, and net strangeness over our realistic hypersurface at $\sqrt{s} = 19.6\gev$  from particles within a rapidity cut of $|y|<1$. The mean value, standard deviation, and higher cumulant ratios are shown in Fig. \ref{fig:conserved_mean_and_fluctuations}. In general, the mean values exhibit a decreasing trend with increasing patch energy and for large patch energies they approach the grand-canonical values. This is because for smaller patch energy particles prefer to be at midrapidity, rather than at high rapidity. The jumps in Fig. \ref{fig:conserved_mean_and_fluctuations} are mainly coming from the way of splitting hypersurface into patches. It becomes evident from Fig. \ref{fig:conserved_mean_and_fluctuations}, where results for two ways of splitting into patches are shown. The difference between them is to be understood as a systematic uncertainty of our approach, which in our case does not exceed 2\% for the energy and baryon number. The standard deviations of conserved quantities (and therefore the scaled variances $\kappa_2/\kappa_1$), quantifying the strength of fluctuations, are systematically smaller for the microcanonical sampler when compared to \texttt{iSS}. One can further observe a weak trend that the higher the patch energy is, the more these quantities fluctuate. In addition, for most choices of patch energy, both skewness and kurtosis are consistent with the zero, which is the expectation of normal distribution. It is not clear, if this property is connected to our assignment of integer charges to the patches, or it is a physical effect. Scaled skewness and kurtosis are different from zero for the grand-canonical iSS sampler.

\subsection{Transverse momentum spectra and flow}

\begin{figure*}
    \centering
    \includegraphics[width=0.45\textwidth]{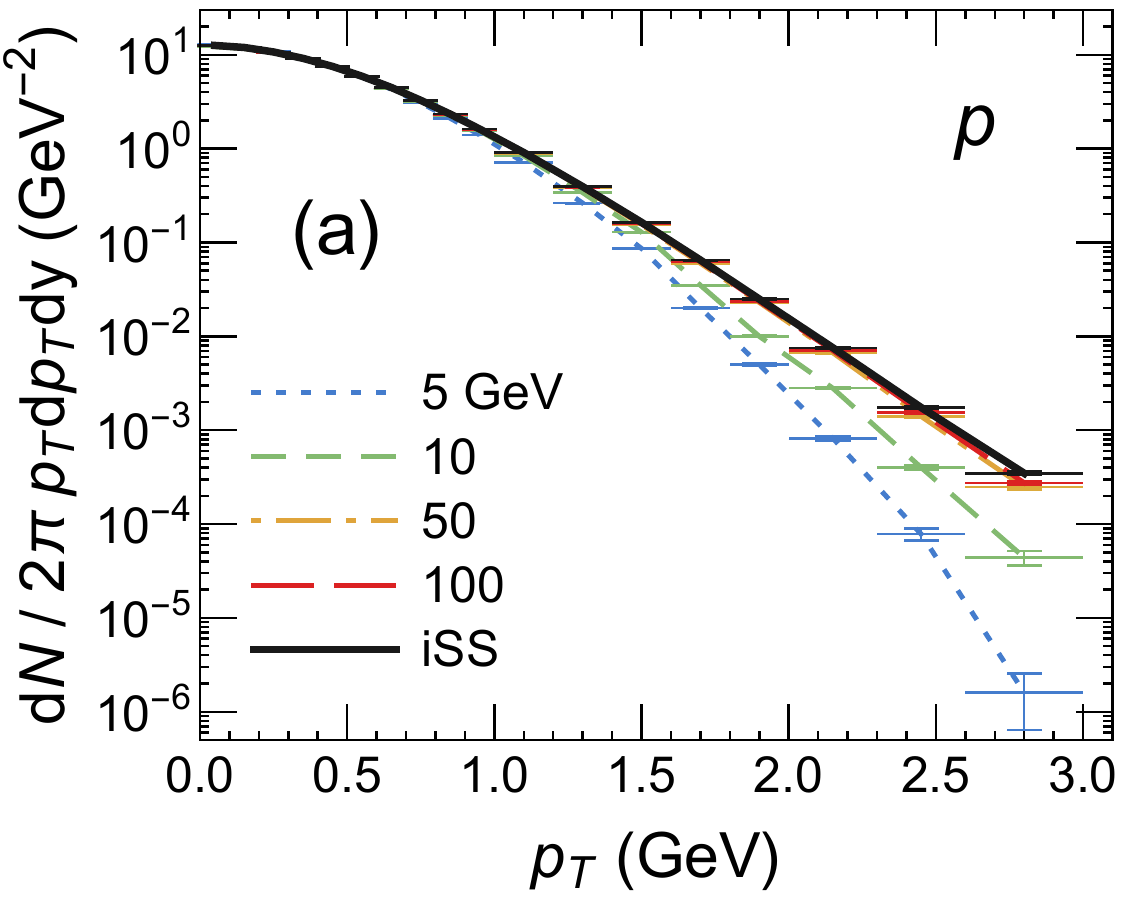}
    \includegraphics[width=0.45\textwidth]{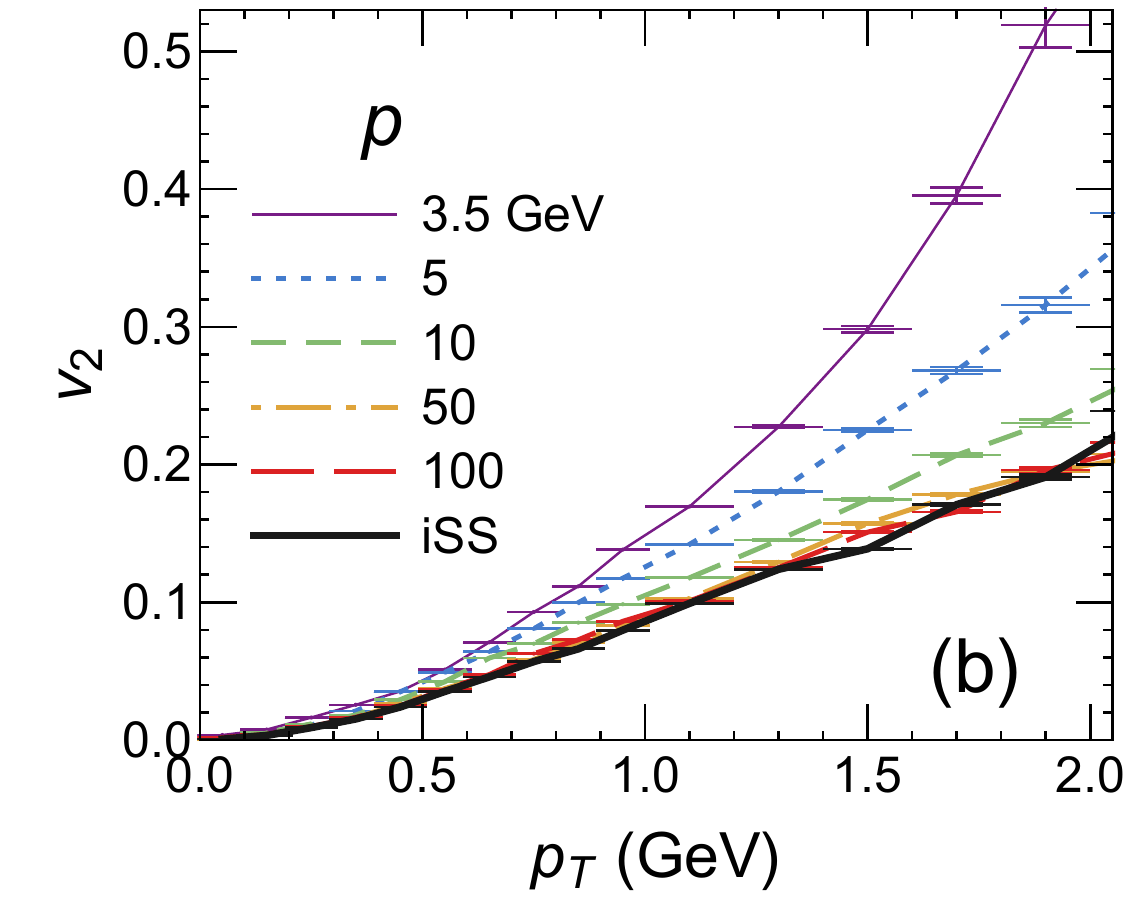}
    \caption{Momentum spectra (left) and elliptic flow (right) of protons depending on the patch energy. At large enough patch size both approach the grand-canonical limit (iSS sampler). The patch splitting algorithm used was: largest energy cell and $\Delta r^2/d_0^2 + (\Delta T / \sigma_T)^2 + (\Delta \mu_B / \sigma_{\mu_B})^2$ distance.}
    \label{fig:spectra_flow}
\end{figure*}

Mean multiplicity, correlations, and fluctuations are affected mostly by baryon number, strangeness, and charge conservation. Energy and momentum conservation play a much smaller role there. The only exception is the total number of particles, which is sensitive to energy conservation. However, energy and momentum conservation influence other observables, such as momentum spectra and correlations. The effects are qualitatively very similar for pions, kaons, and protons. Therefore, in Fig. \ref{fig:spectra_flow} we show only protons.

Transverse momentum distributions are expectedly suppressed at high momenta due to energy conservation. Indeed, it is clear that a patch of total energy of 5 GeV should on average contain less protons with transverse momentum $p_T = 3$ GeV than a patch of 10 GeV. However, at much smaller momenta than the patch energy the microcanonical distributions approach the grand-canonical ones. At high momenta microcanonical sampling always results in a cutoff due to the limited total energy in a patch. This is unlike the grand-canonical Boltzmann distribution, which has non-zero probability for arbitrarily high momenta, since it assumes the presence of a heat bath.

Reproduction of the grand-canonical elliptic flow in the limit of a large patch is a good test that our sampler properly takes into account the local velocities of the cells. The elliptic flow is defined as
\begin{equation}\label{Eq:v2_definition}
v_2 (p_T) \equiv \langle \cos(2\phi_i) \rangle_{p_{T,i}\in \text{\{$p_T$ bin}\}}  \,.
\end{equation}
where $\phi_i$ is the angle with respect to the reaction plane. Elliptic flow is sensitive to the local variations in hydrodynamic cell velocities, temperatures, and chemical potentials within a patch. Our sampling algorithm takes into account these local variations and thus is able to reproduce the flow, as demonstrated in Fig. \ref{fig:spectra_flow}. At smaller patch energies we observe an interesting effect of momentum conservation: for smaller patches $v_2$ is larger at high momenta. We conjecture that this is caused by momentum conservation. In the grand-canonical sampler total momentum of the patch can fluctuate, therefore the momentum anisotropy, which reflect the anisotropy of the collective flow field $\vec{u}$, is smeared out compared to the microcanonical sampler. In the microcanonical ensemble for smaller patches we obtain larger integrated elliptic flow and smaller $p_T$, in other words $\langle v_2^{MCE} \rangle > \langle v_2^{GCE} \rangle$ and $\langle p_T^{MCE} \rangle < \langle p_T^{GCE} \rangle$.

The dependence of $v_2(p_T)$ is used to quantify the viscosity of the quark-gluon plasma, because alarger viscosity to entropy density ratio $\eta/s$ leads lower $v_2(p_T)$. The values of $\eta/s \approx 0.08 - 0.16$ were obtained from fitting experimental data \cite{Niemi:2012ry,Karpenko:2015xea,Bernhard:2019bmu}. However, these works do not account for local microcanonical conservation laws. Our result in Fig. \ref{fig:spectra_flow} demonstrates, that a larger $\eta/s$ may be necessary to reproduce experimental data, if the local microcanonical conservation laws are included.

\subsection{Correlations as a function of pseudorapidity}

\begin{figure*}
    \centering
    \includegraphics[width=0.32\textwidth]{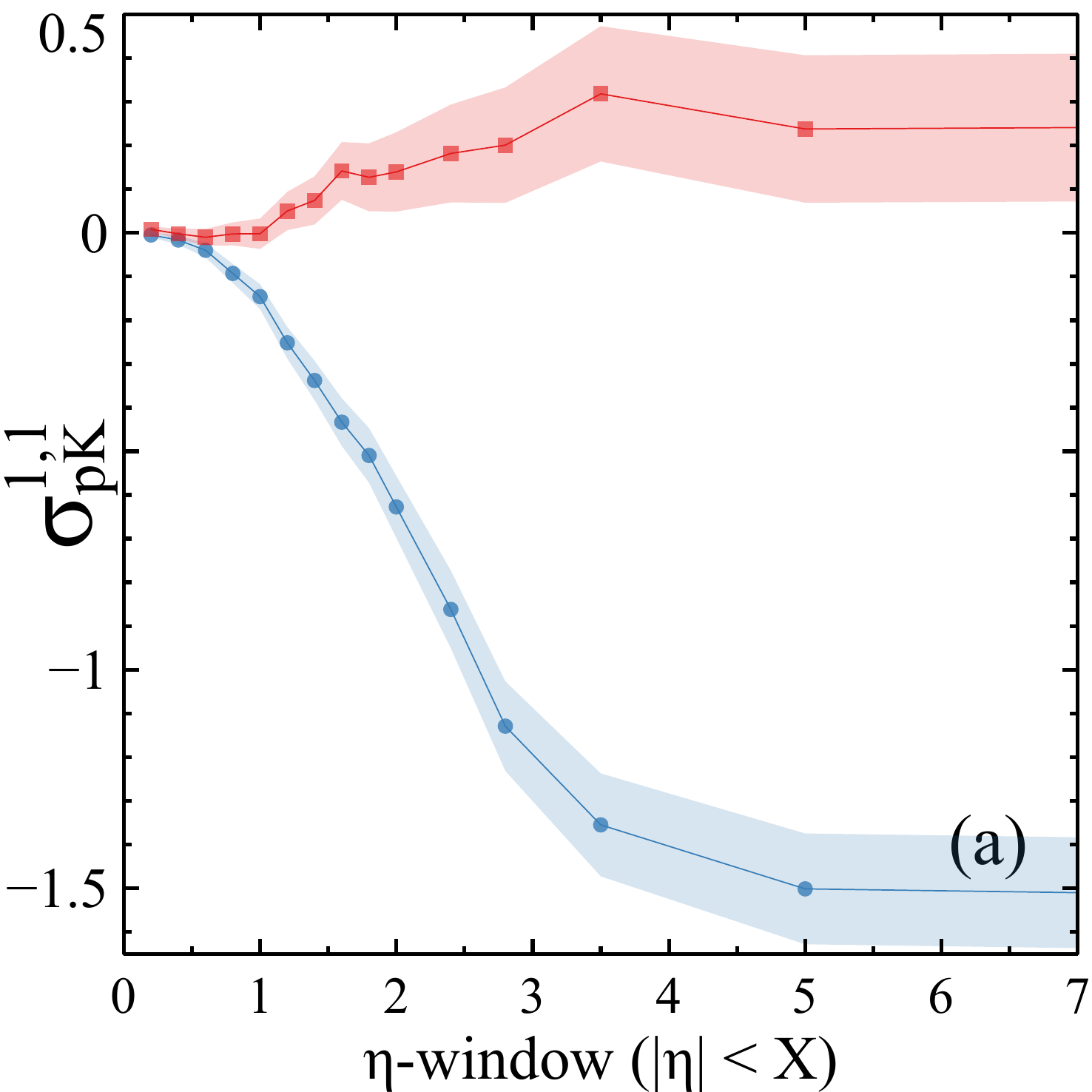}
    \includegraphics[width=0.32\textwidth]{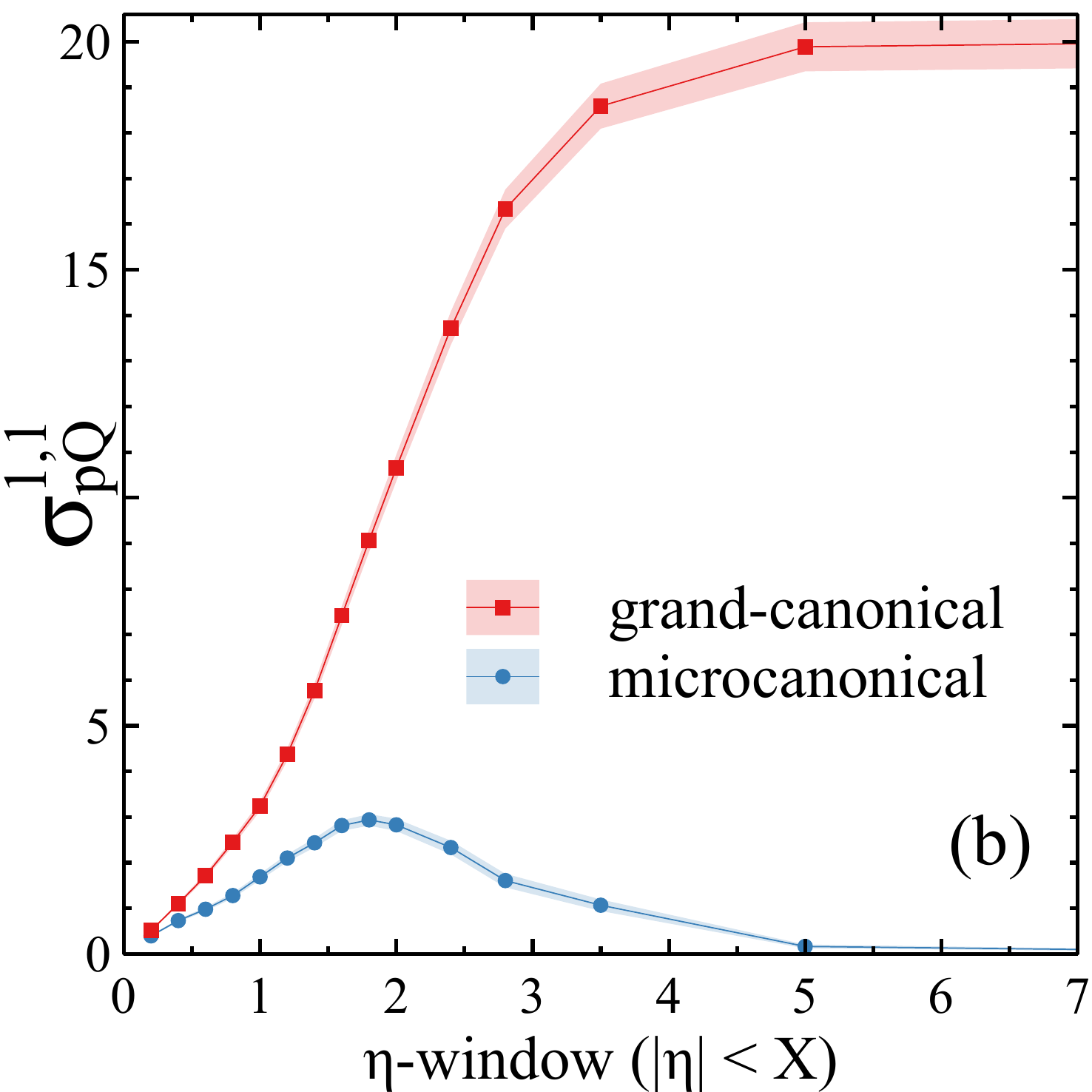}
    \includegraphics[width=0.32\textwidth]{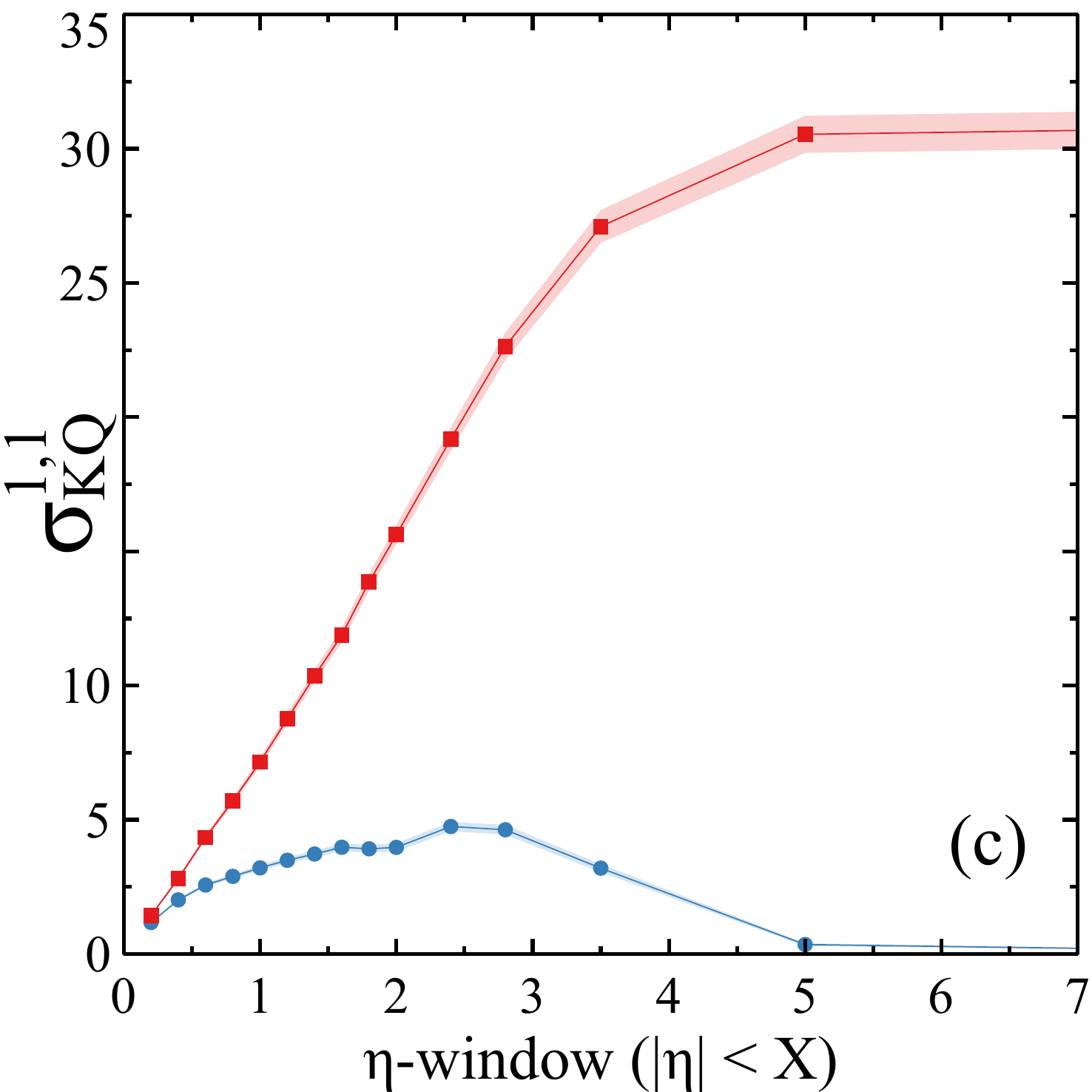}
    \caption{Unnormalized correlations of net proton, net kaon, and net charge from microcanonical and grand-canonical sampler as a function of pseudorapidity gap $|\eta|$. Here we used a patch energy of $E_{patch}=10\gev$. The patch splitting algorithm used was: largest energy cell and $\Delta r^2/d_0^2 + (\Delta T / \sigma_T)^2 + (\Delta \mu_B / \sigma_{\mu_B})^2$ distance. }
    \label{fig:corr_pseudorapidity}
\end{figure*}

Correlations between particle multiplicities are already discussed above as a function of the patch size, mainly to understand the effect of the patch size. For patch rest-frame energies larger than 10 GeV we find that correlations are almost unchanged, see Fig. \ref{fig:corr}. Here we would like to explore a correlation observable, similar to the one measured by the STAR collaboration \cite{Adam:2019xmk}. For this purpose we select a patch rest frame energy to be 10 GeV. STAR measured correlations of net protons ($p - \bar{p}$), net kaons ($K^+ - K^-$), and net charge as a function of a pseudorapidity gap $\eta = \frac{1}{2} \log \frac{|p|+p_z}{|p| - p_z}$. Because STAR published unnormalized correlations, we also do not normalize them here:
\begin{equation}
\sigma^{1,1}_{AB} \equiv \langle (A - \bar{A}) (B - \bar{B})\rangle \,.
\end{equation}
The range of rapidity in the  STAR measurement is limited to $|\eta| < 0.5$, while we can simulate the whole rapidity range. It is important to note, however, that we do not perform any afterburner simulation and do not include resonance decays. Also, our centrality selection is different from \cite{Adam:2019xmk}. Therefore a direct comparison to STAR data is not meaningful. However, we are able to pinpoint the effect of conservation laws. The patch definition used here is the one corresponding to Fig. \ref{fig:different_splitting} (d): starting from largest energy cell and clustering cells by distance $\Delta r^2/d_0^2 + (\Delta T / \sigma_T)^2 + (\Delta \mu_B / \sigma_{\mu_B})^2$.

Our comparison of the rapidity-dependent correlations between micro- and grand-canonical samplers is shown in Fig. \ref{fig:corr_pseudorapidity}. The most prominent feature is that at small rapidity the correlation between net proton and net kaon has a negative slope, if the local conservation laws are included. This is similar to the results of \cite{Adam:2019xmk} at all energies higher than 7.7 GeV, and this feature is not reproduced by the grand-canonical sampler. At 7.7 GeV the measured net-$pK$ correlation has a positive slope. We conjecture that the positive slope may originate from the conservation laws, when total net baryon number and total net charge are sufficiently large. Another possibility is resonance decays. We are able to handle these effects separately, thus we are potentially able to identify the reason of the measured positive slope. This task is left for a future work, however.

At small $\eta$ the net-$pQ$ and net-$KQ$ correlations have positive slopes as a function of $|\eta|$ both for micro- and grand-canonical samplers. At large $\eta$ there is a large difference between the samplers. The reason is the following. In the limit of large $\eta$ the charge within $|\eta|$ interval  does not fluctuate in the microcanonical sampler by construction. Therefore in each sample $Q - \langle Q \rangle = 0$ and net-$pQ$ and net-$KQ$ correlations approach zero.

\section{Summary, discussion, and outlook} \label{sec:summary}

We have introduced local microcanonical sampling over the hydrodynamical hypersurface. The localness is reached by splitting the hypersurface into patches --- spatially compact regions, where conservation laws are enforced in every sample using a novel sampling algorithm \cite{Oliinychenko:2019zfk}. This algorithm conserves energy, momentum, baryon number, strangeness, and electric charge in each patch. It also preserves local variations of velocity, temperature, and chemical potential within a patch.

The idea of patches combined with the sampling algorithm allows to study a rich variety of microcanonical effects in heavy ion collisions. We have explored means, fluctuations, and correlations of multiplicity distribution for pions, koaons, and protons within a rapidity cut; means and fluctuations of conserved charges within a rapidity cut; transverse momentum spectra and flow; and correlations of net protons, kaons, and charge as a function of the pseudorapidity gap. All these observables except the last one were studied as a function of the patch size. For the smallest patch size microcanonical effects are the strongest, but many effects, in particular for fluctuations and correlations, do not vanish even in the thermodynamic limit, as expected from analytical calculations \cite{Begun:2004pk}.

While the microcanonical sampling is mathematically rigorous and well-defined, the patch splitting procedure and its parameter, the patch size choice are to a large extent a matter of choice. Which procedure and which patch size should one select to simulate heavy ion collisions? The variation of the patch splitting algorithm changes our results, but not too much; we consider these changes shown in Figs. \ref{fig:different_splitting}, \ref{fig:mean_scaledvar_particle}, \ref{fig:corr}, and \ref{fig:conserved_mean_and_fluctuations} as a systematic uncertainty of our method. The patch size, in contrast, plays an important role for every observable studied, as long as this size is not too large. Above the patch rest frame energy of around 10 GeV observables depend only very slightly on it. This makes it tempting to choose 10 GeV as a ``reasonable'' patch size. However, we would like to underline that the question about correct patch size is not algorithmic, but physical. The quark-gluon plasma created in heavy ion collisions has a surface tension (which is usually neglected in hydrodynamic simulations, except \cite{Steinheimer:2012gc}) and, therefore, droplets may be formed. When these droplets hadronize they form droplets of hadron gas. In this case the right patch size should be equal to the size of the droplet, and the droplet scenario can be identified by the microcanonical effects we have listed and explored: high-momentum suppression, $v_2(p_T)$ enhancement at high $p_T$, stronger suppression of fluctuations, and enhancement of correlations. The latter has already been pointed out in \cite{Koch:2005pk} in this context of droplet formation due to spinodal instabilities. Although in principle these effects are always present, they can be observed easily only if the droplet energy is of order of 10 GeV or less. The high-$p_T$ suppression should be the most susceptible to experimental observation.
Here we considered droplets uniform in size. This was done to study the microcanonical effects systematically, but in general there is no reason to assume that droplets have the same size. Qualitatively, the microcanonical effects we have observed, should also occur if the droplets have different sizes.

Our present results cannot be directly compared to experimental data. Here we have purposely only  studied the properties and effects of the local microcanonical particlization in isolation, without subsequent resonance decays or hadronic afterburner. This allows to understand the sampler and its systematics better, before using it in a larger framework. Now, that the sampler is explored and tested, it can be used as a part of the hybrid (hydrodynamics + transport) approach. For example, it would be interesting to see if it can reproduce the net-p, net-K, and net charge correlations measured recently by the STAR collaboration \cite{Adam:2019xmk}. Also, effects of conservation laws on observables related to the chiral magnetic effect, as well as small systems should be studied and be compared to the data. It is important to notice that in case of small enough patches ($E_{patch} < 10$ GeV) the ratios of microcanonical equilibrium hadron yields are different from the grand-canonical ones. Therefore, the particlization criterion with microcanonical particlization has to be refitted to match the data. This is a known fact in the thermal models: in small systems, such as pp or pPb collisions, the temperature describing hadron yields is different in the microcanonical ensemble compared to the grand-canonical one \cite{Becattini:2004rq}.

Furthermore, the presented sampling algorithm allows a consistent particlization of fluctuating hydrodynamics. Therefore, it can be applied to study the physics of critical point in heavy ion collisions. Certain improvements of the sampler may be of interest, such as introducing viscous corrections and quantum statistics. Finally, it turns out that, contrary to the usual grand-canonical particlization our approach allows for sampling particles with negative individual weights. As discussed, although the weight of the whole multi-particle sample has to be positive, contributions from individual particles do not need to be. This feature may allow to tackle the problem of negative Cooper-Frye contributions from a new perspective. However, we leave this idea for future studies. All our tests are intentionally performed for hypersurfaces with negligible negative contributions.

\begin{acknowledgments}
D.O. would like to thank Prof. S. Pratt for his constructive criticism of the sampling. This work was supported by the U.S. Department of Energy, Office of Science, Office of Nuclear Physics, under contract number DE-AC02-05CH11231 and received support within the framework of the Beam Energy Scan Theory (BEST) Topical Collaboration. S.S. is grateful to the Natural Sciences and Engineering Research Council of Canada.

Computations by S.S. were made on the supercomputer Bel\'uga, managed by Calcul Qu\'ebec and Compute Canada. Computational resources used by D.O. were provided by Center for Scientific Computing in Frankfurt.
\end{acknowledgments}

\appendix

\section{Special case --- microcanonical gas of relativistic massless particles} \label{appendix:massless_microcanonical}

The simplest test of our sampler is against the analytically known case of a classical microcanonical relativistic massless gas. Here by microcanonical we mean that all possible phase space cells $(\vec{x}, \vec{p})$ can be occupied with equal probability, but the total energy $E$ is conserved and the total momentum is zero. The number of particles $N$ is allowed to vary. Then the probability to have $N$ particles with momenta $\{p_i\}_{i=1}^{N}$ is

\begin{eqnarray} \label{Eq:A1}
w_N^{mce}(\{p_i\}) \sim \frac{V^N}{(2\pi \hbar)^{3N}} \frac{1}{N!} \prod_{i=1}^N d^3p_i  \delta_{(E - \sum p_i)} \, \delta^{(3)}_{(\sum \vec{p}_i)}
\end{eqnarray}
where $V$ the volume of the system. In terms of our sampler, this is a special case, where a patch consists of just one static ($u^{\mu} = (1, 0,0,0)$) cell with $d\sigma_{\mu} = (V, 0,0,0)$.The momenta in Eq. (\ref{Eq:A1}) can be integrated out analytically, which provides the following distribution of the total number of particles~\cite{Milburn:1955zz,Werner:1995mx}:

\begin{eqnarray} \label{Eq:A2}
w_N^{mce} \sim \frac{(VE^3)^N}{(16 \pi^2 \hbar^3)^{N}} \frac{(4N-4)! }{N!(2N-1)! (2N - 2)! (3N-4)!}
\end{eqnarray}

It is convenient to rewrite this distribution in terms of the grand-canonical mean for the same system. The grand-canonical probability is

\begin{eqnarray} \label{Eq:A3}
w_N^{gce}(\{p_i\}) \sim \frac{V^N}{(2\pi \hbar)^{3N}} \frac{1}{N!} \prod_{i=1}^N d^3p_i \, e^{- p_i/T}
\end{eqnarray}

Integrating out momenta, one obtains the Poisson distribution $w_N^{gce} \sim \bar{N}_{gce}^N / N! $, its mean being $\bar{N}_{gce} = \frac{V T^3}{\pi^2 \hbar^3}$. The mean energy per particle is computed from Eq. (\ref{Eq:A3}), $E/\bar{N}_{gce} = 3T$. Eliminating the temperature $T$, one obtains

\begin{eqnarray} \label{Eq:A4}
\bar{N}_{gce}^4 = \frac{VE^3}{27 \pi^2 \hbar^3}
\end{eqnarray}

Therefore, one can express the microcanonical particle number distribution via the grand-canonical average at the same average total energy and volume:

\begin{eqnarray} \label{Eq:A5}
  w_N \sim \left(\frac{27}{16} \bar{N}_{gce}^4 \right)^N \frac{(4N-4)!}{N! (2N-1)! (2N - 2)! (3N-4)!}  
\end{eqnarray}

 The cumulants $\kappa_i$ of distribution (\ref{Eq:A5}) can be computed analytically as $\kappa_i = (\partial^i F /\partial t^i) |_{t=0}$, where $F(t)$ is the cumulant generating function:

\begin{eqnarray}
  F(t) = \log \sum_{N=2}^{\infty} w_N e^{tN} = \log c + 2t + \\ \nonumber \log \,\, _2F_5 \left(\{\frac{5}{4},\frac{7}{4} \}, \{\frac{4}{3}, \frac{5}{3}, 2, \frac{5}{3}, 3 \}, e^t \bar{N}_{gce}^4 \right) \,,
\end{eqnarray}

Here $_pF_q$ is a generalized hypergeometric function and $c$ is a constant. The exact expressions for the cumulants $\kappa_{1-4}$ are, therefore, available analytically, but we do not provide them here, because they are bulky and hardly informative. Instead, we show the expansions of certain combinations in the thermodynamic limit $\bar{N}_{gce} \to \infty$, which are more interesting and instructive:

\begin{eqnarray}
  \bar{N} \equiv \kappa_1 &=& \bar{N}_{gce} + \frac{1}{2} + \frac{65}{288} \bar{N}_{gce}^{-1} + O(\bar{N}_{gce}^{-2}) \\
  \kappa_2/\kappa_1 &=& \frac{1}{4} - \frac{1}{8} \bar{N}_{gce}^{-1} + O(\bar{N}_{gce}^{-2}) \\
  \kappa_3/\kappa_2 &=& \frac{1}{4} + \frac{1865}{5184}  \bar{N}_{gce}^{-2} + O(\bar{N}_{gce}^{-3}) \\
  \kappa_4/\kappa_2 &=& \frac{1}{16} + \frac{5}{81} \bar{N}_{gce}^{-2} + O(\bar{N}_{gce}^{-3}) 
\end{eqnarray}

From this one can see that in the thermodynamic limit the microcanonical mean number of particles is larger by $\frac{1}{2}$ than the grand-canonical one. This counter-intuitive result does not contradict the theorem about ensemble equivalence, because $\bar{N} / \bar{N}_{gce} \to 1$ is still fulfilled at $\bar{N}_{gce} \to \infty$. A non-zero difference between microcanonical and grand-canonical yields was also reported when only energy conservation (but not momentum) is taken into account, see Eq. (9) of \cite{Begun:2004pk}. The scaled variance $\kappa_2/\kappa_1$, and the ratios $\kappa_3/\kappa_2$, and  $\kappa_4/\kappa_2$ are not 1, like in the grand-canonical case, but rather $\frac{1}{4}$, $\frac{1}{4}$, and $\frac{1}{16}$.

We are interested in comparing the cumulants of the distribution (\ref{Eq:A5}) to the cumulants of the distribution produced by our sampler. As the distribution is sampled indirectly, using random $2\leftrightarrow 3$ Metropolis steps in momentum space, it is non-trivial, that the distribution in Eq. (\ref{Eq:A5}) is reproduced. We found this example a very useful testbed for our algorthm. Any error in the proposal or acceptance probabilities (Eq. \ref{Eq:sampling_acceptance_probability}) dramatically changes all moments, including the mean.

\begin{figure}
    \centering
    \includegraphics[width=0.49\textwidth]{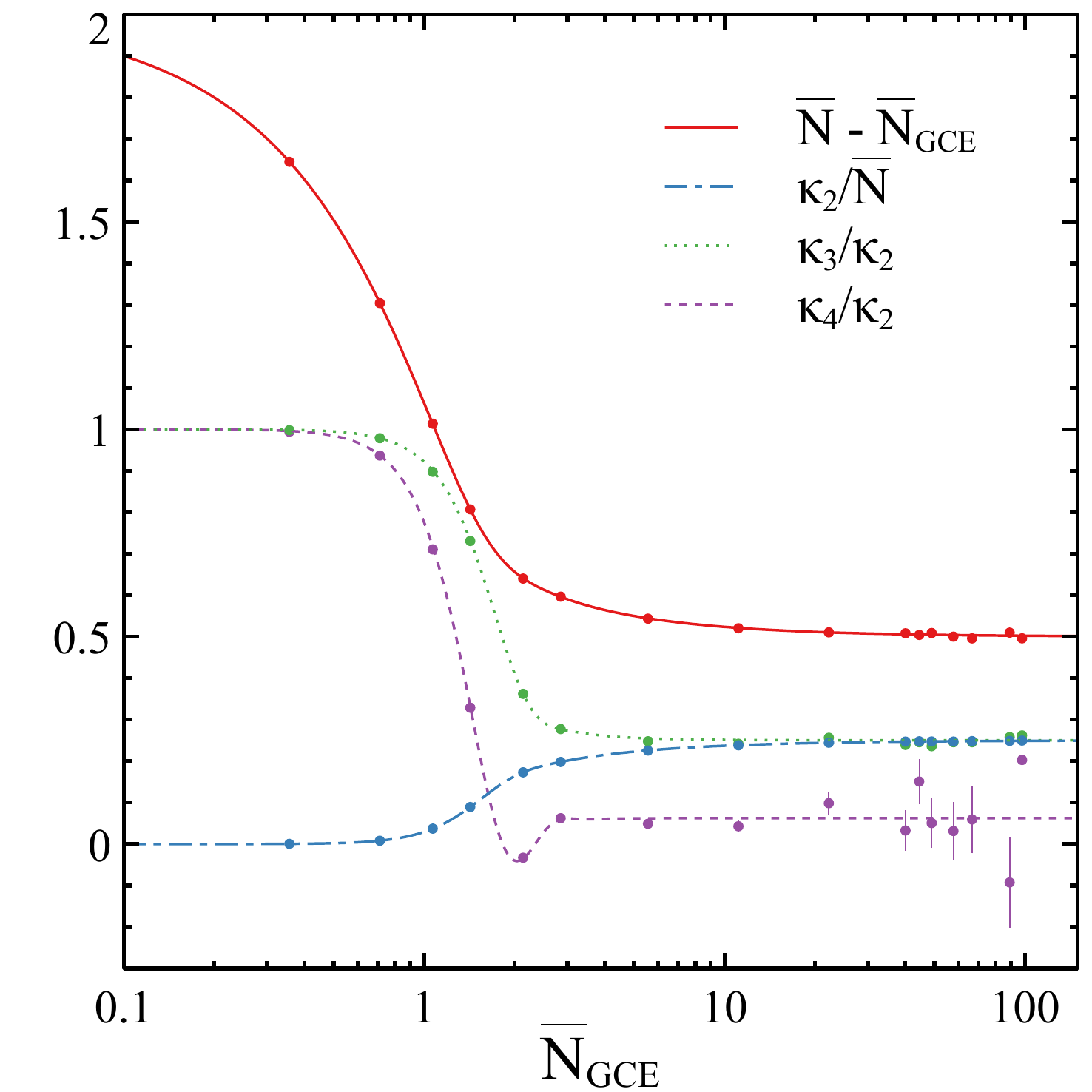}
    \caption{Demonstration that the analytically computed statistics of massless microcanonical relativistic gas (lines) agree with the ones generated by the sampler (circles).}
    \label{fig:A1}
\end{figure}

In Fig. \ref{fig:A1} we demonstrate that the mean, scaled variance, and cumulant ratios $\kappa_3/\kappa_2$ and  $\kappa_4/\kappa_2$ of the generated distribution agree with the analytical results computed from Eq. (\ref{Eq:A5}). For each point $N_{ev} = 10^6$ samples were generated. Statistical errors were estimated following \cite{Luo:2011tp} as
\begin{eqnarray}
  \Delta^2(\bar{N}) &=& \sigma^2 / N_{ev} \\
  \Delta^2(\kappa_2/\kappa_1) &=& (\mu_4 - \sigma^4) / N_{ev} \\
  \Delta^2(\kappa_3/\kappa_2) &=& 6 \sigma^2 / N_{ev} \\
  \Delta^2(\kappa_4/\kappa_2) &=& 24 \sigma^4 / N_{ev}
\end{eqnarray}

 Here $\mu_4$ is the fourth central moment, $\mu_4 = \sum_{i=1}^{N_{ev}} (N_i - \bar{N})^4 / N_{ev}$, and $\sigma$ is the variance, $\sigma^2 = \sum_{i=1}^{N_{ev}} (N_i - \bar{N})^2 / N_{ev}$. The equations for $\Delta^2(\kappa_3/\kappa_2)$ and  $\Delta^2(\kappa_4/\kappa_2)$ are simplified error estimates derived assuming a Gaussian distribution.
 
\section{Special case II --- microcanonical sampling of hadron resonance gas} \label{appendix:Begun_microcanonical}

\begin{figure}
    \centering
    \includegraphics[width=0.5\textwidth]{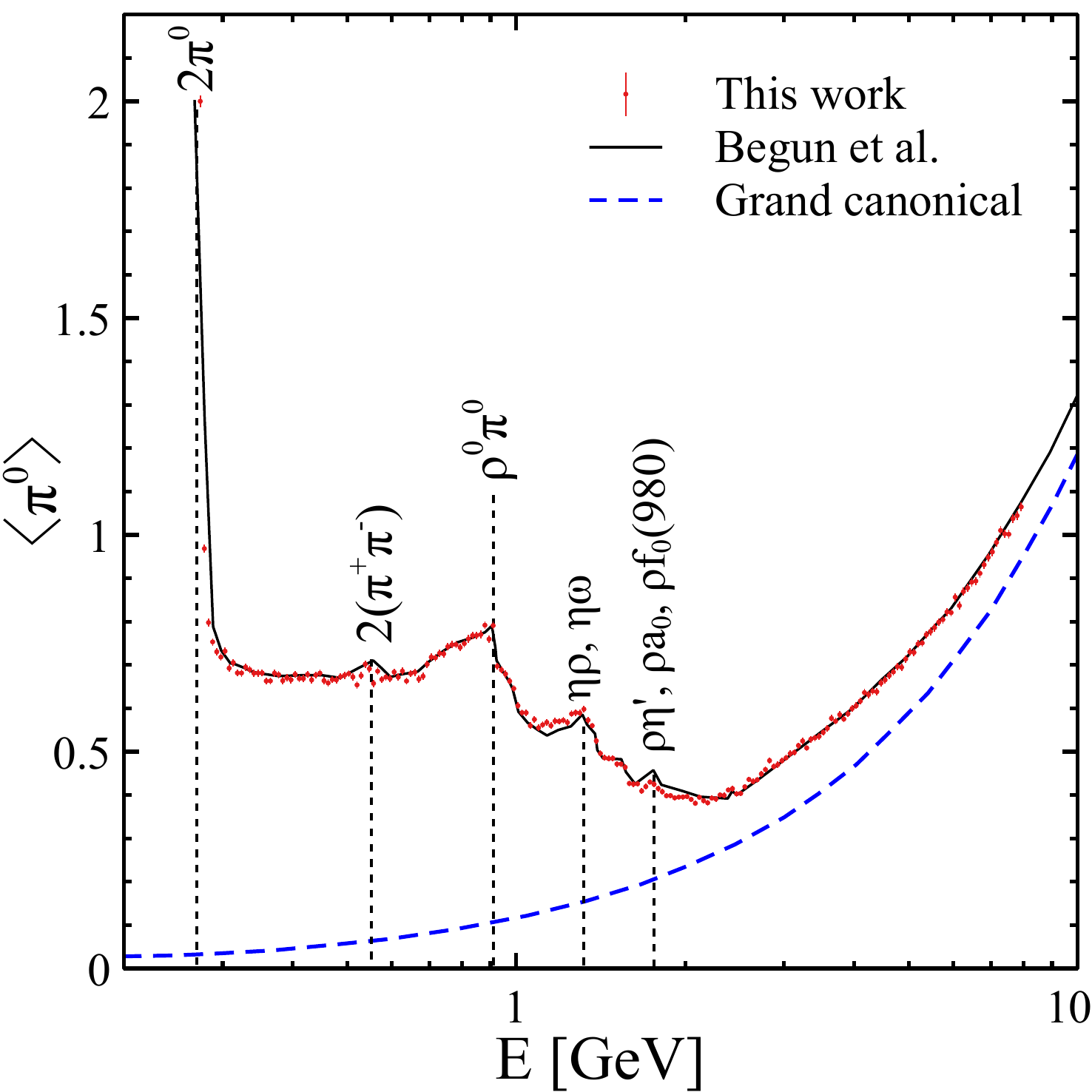}
    \caption{Test of our sampler we where we microcanonically sample a hadron gas in a single, static cell with zero net baryon number, strangeness, and charge. Here we show a comparison of the threshold effects in the  $\pi^0$ yield with the results of presented in Fig. 12 of Begun et al \cite{Begun:2005qd}.}
    \label{fig:B1}
\end{figure}

While the microcanonical sampling of one massless hadron species in Appendix \ref{appendix:massless_microcanonical} is a sensitive test of the sampling algorithm, it does not check the aspects of the sampling related to channel selection. Neither does it test baryon number, strangeness, and charge conservation. To test the latter, in this section we apply our sampler for a special case where a patch consists of just one static cell with $d\sigma_{\mu} = (V, 0,0,0)$, as in Appendix \ref{appendix:massless_microcanonical}. However, multiple hadronic species are allowed. Multiplicity distributions are not calculable analytically in this case, but they were thoroughly studied using Monte-Carlo sampling \cite{Werner:1995mx,Becattini:2004rq,Begun:2005qd}. The microcanonical sampler used in \cite{Werner:1995mx} is based on Metropolis algorithm, a faster sampler \cite{Becattini:2004rq} uses importance sampling with canonical distribution as an envelope. Unlike our sampler, both require direct computation of the microcanonical partition function.

Here we test that our sampler reproduces the non-trivial threshold effects on the $\pi^0$ yield, which were shown in Fig. 12 of \cite{Begun:2005qd}. In Fig. \ref{fig:B1} we demonstrate a good agreement with a previous calculation, which used a dedicated microcanonical sampler. Minor discrepancies might be attributed to possible differences in hadron tables, such as the mass of $f_0$ meson and different number of mesonic resonances between 1 and 2 GeV. In our calculation we use the default particle table of SMASH transport code \cite{Weil:2016zrk} with the $\pi^0$ mass set to $135$ MeV, and the  $\pi^{\pm}$ mass set to 138 MeV.

In Fig. \ref{fig:B1} one can also see interesting physical effects which were already studied in \cite{Begun:2005qd}. The minimal amount of particles in the microcanonical case is two, because one or zero particles cannot fulfil energy and momentum conservation. Therefore, at the smallest energy only a state with $2\pi^0$, the lightest hadron, is accessible. At slightly higher energy the $\pi^+\pi^-$ state opens and the $\langle \pi^0 \rangle$ yield drops down to $2/3$. Then, with increasing energy $\langle \pi^0 \rangle$ grows until the energy crosses a threshold to form a new state. Some of such thresholds are marked explicitly in Fig. \ref{fig:B1}.

At energies above $3$ GeV threshold effects are not pronounced anymore, although in principle new many-particle thresholds continue to open at arbitrarily high energies. Nevertheless, even at $E = 10$~GeV the microcanonical average still differs from the grand-canonical average at the same average energy. Their ratio approaches 1 in the thermodynamic limit, but a finite difference remains. This is similar to the microcanonical massless case, studied above (Appendix \ref{appendix:massless_microcanonical}), however the  finite difference is smaller than $1/2$.

\section{Integrals related to hydro hypersurface} \label{appendix:integration}

To compute the total energy and charges, Eq. (\ref{Eq:cons_laws}), one needs to compute integrals over the Boltzmann distribution. For completeness we list the expressions for these integrals here. The integrals in Eq. (\ref{Eq:cons_laws}) are Lorentz-invariant and most comfortable to compute in the frame, where $u = (1,0,0,0)$. In this frame the hypersurface normal is
\begin{equation}
    d\sigma_{\mu} \xrightarrow{u^{\mu}} d\sigma'_{\mu} = \Lambda_{\mu}^{\nu} d\sigma_{\nu} \,,
\end{equation}
where $\Lambda_{\mu}^{\nu}$ is a boost matrix. It follows from the explicit form of $\Lambda_{\mu}^{\nu}$ that $d\sigma'_0 = u^{\mu}d\sigma_{\mu}$. Consequently the integral for the density coincides with a known expression in literature (e.g. Appendix of \cite{Denicol:2018wdp}). Noticing that integrals over any odd function vanish from $-\infty$ to $\infty$, and denoting $z\equiv m/T$, one obtains in this frame

\begin{eqnarray} \label{eq:cf_integrals}
\int p^{\nu} d\sigma_{\nu} e^{- p^{\alpha}u_{\alpha}/T} \frac{d^3p}{p^0} =
4\pi d\sigma'_0 \int_0^{\infty} p^2 e^{-p^0/T} dp \nonumber\\ = 4\pi d\sigma'_0 T^3 z^2 K_2(z) \\
\int p^{\mu} p^{\nu} d\sigma_{\nu} e^{- p^{\alpha}u_{\alpha}/T} \frac{d^3p}{p^0} = 
4\pi \nonumber \\ \left(d\sigma'_0 \int_0^{\infty} p^0 e^{-p^0/T} p^2 dp ,\,
d\vec{\sigma}' \frac{1}{3} \int_0^{\infty} p^2/p^0 e^{-p^0/T} p^2 dp \right) =\\
= 4\pi T^4 z^2 \left( d\sigma'_0 (3K_2(z) + z K_1(z)), \, d\vec{\sigma}' K_2(z) \right)
\end{eqnarray}

Here $d\vec{\sigma}' \equiv (d\sigma'_1,d\sigma'_2,d\sigma'_3)$. After computing these integrals as shown above, one has to boost the 4-momentum back to the computational frame.
Formulas for quantum statistics can be obtained by adding $\sum_{k=1}^{\infty} (\pm 1)^{k+1} e^{\mu k/T}$ in front of the expressions and substituting $T \to T/k$, $z \to z k$. Here the  $+$ sign is for bosons and  the $-$ is for fermions.

\section{Phase space sampling and integrals}

Here we present the relevant expressions for the two- and three-body phase space integrals $R_2$ and $R_3$. While the analytical expression for $R_2$ is well-known \cite{Seifert:2017oyb}, $R_3$ is less common, although available in the literature. The definition of a phase space element for $n$ dimensions is

\begin{eqnarray}
dR_n(\sqrt{s}, m_1, m_2, \dots, m_n) = (2\pi)^4 \frac{1}{(2\pi)^{3n}}  \nonumber \\
\frac{d^3p_1}{2 E_1} \frac{d^3p_2}{2 E_2} \dots \frac{d^3p_n}{2 E_n} \delta^{(4)}(P_{tot}^{\mu} - \sum P_i^{\mu}) \,,
\end{eqnarray}

where $P_{tot}^{\mu} P^{tot}_{\mu} = s$. The whole expression is Lorentz invariant and can be evaluated in any convenient frame. Evaluating $R_2$ in the center of mass frame one finds:

\begin{eqnarray}
dR_2(\sqrt{s}, m_1, m_2) = \frac{1}{(2\pi)^2} \frac{d^3p_{CM}}{4 \sqrt{p_{CM}^2 + m_1^2} \sqrt{p_{CM}^2 + m_2^2}} \nonumber\\
\delta(\sqrt{s} - \sqrt{p_{CM}^2 + m_1^2} - \sqrt{p_{CM}^2 + m_2^2}) \\
\delta(\sqrt{s} - \sqrt{p_{CM}^2 + m_1^2} - \sqrt{p_{CM}^2 + m_2^2}) =\nonumber\\ \frac{E_1 E_2}{p_{CM} \sqrt{s} } \, \delta(p_{CM} - p_{CM}(\sqrt{s}, m_1, m_2)) \\
dR_2(\sqrt{s}, m_1, m_2) = \frac{d^2\Omega_{CM}}{4\pi}  \frac{p_{CM}(\sqrt{s}, m_1, m_2)}{4 \pi \sqrt{s}} \,,
\end{eqnarray}

where the center of mass momentum, $p_{CM}(\sqrt{s}, m_1, m_2)$, is given by

\begin{eqnarray}
p^2_{CM}(\sqrt{s}, m_1, m_2) = \frac{(s - (m_1 + m_2)^2) (s - (m_1 - m_2)^2)}{4 s}
\end{eqnarray}

Therefore, sampling two-body phase space means just sampling two angles uniformly on a unit sphere in the center of mass frame, $dR_2 / R_2 = d^2\Omega_{CM}/4\pi$ and $R_2 = \frac{p_{CM}}{4 \pi \sqrt{s}}$. To compute $R_3$ one inserts the following identities into the integration:
\begin{eqnarray}
\int \delta(E_{12} - E_1 - E_2) \delta^{(3)}(\vec{p}_1 + \vec{p}_2 - \vec{p}_{12}) d^4p_{12} = 1 \\
\int \delta(E_{12}^2 - \vec{p}_{12}^2 = M_{12}^2) \, dM_{12}^2 = 1
\end{eqnarray}

Then after rearranging the product
\begin{eqnarray}
dR_3 = \frac{(2\pi)^3}{(2\pi)^4} dR_2(M_{12},m_1,m_2) \, dR_2(\sqrt{s}, M_{12}, m_3) \, d M_{12}^2 \\
    = \frac{1}{16 \pi^3 \sqrt{s}} p_{CM}(\sqrt{s},M_{12}, m_3) \, p_{CM}(M_{12},m_1, m_2)  \times \nonumber \\ dM_{12} \frac{d^2\Omega_{12}}{4\pi} \frac{d^2\Omega_{123}}{4\pi}
\end{eqnarray}

This expression provides a simple algorithm to sample 3-body phase space:
\begin{itemize}
    \item Sample $M_{12}$ uniformly in $[m_1+m_2, \sqrt{s} - m_3]$, then accept with probability
          $\frac{p_{CM}(\sqrt{s},M_{12}, m_3) \, p_{CM}(M_{12},m_1, m_2)}{p_{CM}(\sqrt{s}, m_1 + m_2, m_3) \, p_{CM}(\sqrt{s} - m_3,m_1, m_2)}$.
          Repeat until $M_{12}$ is accepted. The acceptance expression uses the fact that $p_{CM}$ is an increasing function of the first argument
          and decreasing function of the second argument.
    \item For particle with mass $m_3$ sample the direction of momentum uniformly in $4\pi$. The magnitude of momentum is $p_{CM}(\sqrt{s},M_{12}, m_3)$
    \item Sample the two-body phase-space for $m_1$ and $m_2$, then boost their momenta to have total momentum $p_{CM}(\sqrt{s},M_{12}, m_3)$
\end{itemize}

The expression above can be numerically integrated to obtain $R_3$, but it turns out that an analytical formula for $R_3$ exists (see Eqs. 54-58 of \cite{Bauberger:1994nk}), which is faster and more reliable. First let us transform $R_3$ to

\begin{eqnarray}
R_3(\sqrt{s},m_1,m_2,m_3) = \frac{1}{128 \pi^3 s} \times \nonumber\\ \int_{x_2}^{x_3} ((t-x_1)(t-x_2)(t-x_3)(t-x_4))^{1/2}\frac{dt}{t} \,,
\end{eqnarray}
where $x_1 = (m_1 - m_2)^2$, $x_2 = (m_1 + m_2)^2$, $x_3 = (m_3 - \sqrt{s})^2$, $x_4 = (m_3 + \sqrt{s})^2$. Then $R_3$ can be expressed via the complete elliptic integrals $\mathrm{K}$, $\mathrm{E}$, $\mathrm{\Pi}$:

\begin{eqnarray}
R_3 = \frac{1}{128 \pi^3 s} \left[
c_{1} \mathrm{K}(\kappa) \right.
+ c_{2} \mathrm{E}(\kappa)
+ c_{3} \Pi\left(\frac{q_{-+}}{q_{--}} \kappa\right) \nonumber \\
\left.+c_{4} \Pi\left(\frac{(m_1-m_2)^2}{(m_1+m_2)^2} \frac{q_{-+}}{q_{--}} \kappa \right) \right] \nonumber \\
\times \Theta(s - (m_1+m_2+m_3)^2)\\
q_{\pm \pm}:=(\sqrt{s} \pm m_{3})^{2}-(m_{1} \pm m_{2})^{2} \\
c_{1}= 4 m_{1} m_{2} \sqrt{\frac{q_{--}}{q_{++}}} \nonumber \\
\times\left\{\left(\sqrt{s}+m_{3}\right)^{2}-m_{3} \sqrt{s}+m_{1} m_{2}\right\} \\
c_{2}=\frac{m_{1}^{2}+m_{2}^{2}+m_{3}^{2}+s}{2} \sqrt{q_{++} q_{--}} \\
c_{3}= \frac{8 m_{1} m_{2}}{\sqrt{q_{++} q_{--}}}\left\{\left(m_{1}^{2}+m_{2}^{2}\right)\left(s+m_{3}^{2}\right)\right.\\ \left.-2 m_{1}^{2} m_{2}^{2}-2 m_{3}^{2} s\right\} \\
c_{4}=-\frac{8 m_{1} m_{2}\left(s-m_{3}^{2}\right)^{2}}{\sqrt{q_{++} q_{--}}} \\
\kappa^{2}=\frac{q_{+-}q_{-+}}{q_{++}q_{--}}
\end{eqnarray}

In case of massless particles expressions for $R_2$ and $R_3$ simplify considerably:

\begin{eqnarray}
R_2(s, 0, 0, 0) = \frac{1}{8\pi} \\
R_3(s, 0, 0, 0) = \frac{s}{256 \pi^3}
\end{eqnarray}

\end{document}